\documentclass[amsmath,amssymb,twocolumn]{revtex4}

\usepackage{mathrsfs}
\usepackage{mathtools}
\usepackage{xfrac}
\usepackage{graphicx}
\usepackage{dcolumn}
\usepackage{bm}
\usepackage{color}
\usepackage{tabularx}
\usepackage{amsmath}

\newcommand{\beq}{\begin{eqnarray}}
\newcommand{\eeq}{\end{eqnarray}}
\newcommand{\bea}{\begin{align}}
\newcommand{\eea}{\end{align}}
\newcommand{\beqq}{\begin{eqnarray*}}
\newcommand{\eeqq}{\end{eqnarray*}}

\newcommand{\trace}[1]{\text{Tr}\,[#1 ]  }

\newcommand{\order}[1]{\langle #1 \rangle }

\newcommand{\up}{\uparrow}
\newcommand{\down}{\downarrow}

\newcommand{\kf}{k_F}

\begin{document}

\title{Odd-parity superconductors with two-component order parameters:  \\ nematic and chiral, full gap and Majorana node}
\author{J\"orn W. F. Venderbos}
\affiliation{Department of Physics, Massachusetts Institute of Technology,
Cambridge, MA 02139, USA}

\author{Vladyslav Kozii}
\affiliation{Department of Physics, Massachusetts Institute of Technology,
Cambridge, MA 02139, USA}

\author{Liang Fu}
\affiliation{Department of Physics, Massachusetts Institute of Technology,
Cambridge, MA 02139, USA}

\begin{abstract}
Motivated by the recent experiment indicating that superconductivity in the doped topological insulator  Cu$_x$Bi$_2$Se$_3$ has an odd-parity pairing symmetry with rotational symmetry breaking, we study the general class of odd-parity superconductors with two-component order parameters in trigonal and hexagonal crystal systems. In the presence of strong spin-orbit interaction, we find two possible superconducting phases below $T_c$, a time-reversal-breaking (i.e., chiral) phase and  an anisotropic (i.e., nematic) phase, and determine their relative energetics from the gap function in momentum space. The nematic superconductor generally has a full quasi-particle gap, whereas the chiral superconductor with a three-dimensional  (3D) Fermi surface has point nodes with lifted spin degeneracy, resulting in itinerant Majorana fermions in the bulk and topological Majorana arcs on the surface. 
\end{abstract}


\maketitle

\draft

\vspace{2mm}

{\it Introduction.---} Unconventional superconductors with non-$s$-wave pairing symmetry have always been an tantalizing topic of condensed matter physics \cite{norman}.  For inversion-symmetric materials, superconducting order parameters can be broadly divided into two types depending: even-parity (such as $s$-wave) and odd-parity (such as $p$-wave). 
There is mounting evidence that the heavy fermion compound UPt$_3$ \cite{sauls, upt3} and the transition metal oxide Sr$_2$RuO$_4$ \cite{rice, maeno} are odd-parity superconductors. In spin-rotational-invariant systems, odd-parity superconductivity generally results from spin-triplet pairing, which can be meditated by spin fluctuations in the vicinity of ferromagnetic instability. On the other hand, the mechanisms and properties of odd-parity superconductivity in spin-orbit-coupled systems have been less explored until recently. In the last few years,  a number of theoretical and numerical studies have shown that in the presence of strong spin-orbit interaction, 
odd-parity pairing can be realized in a broad range of materials without proximity to magnetic instabilities \cite{fuberg, nagaosa, wan, brydon, hosur, yanase, kozii, andofu}. 

In particular, Fu and Berg \cite{fuberg}  proposed that the doped topological insulator Cu$_x$Bi$_2$Se$_3$, which is superconducting below $T_c \sim 3.8$K \cite{cava, ando} and has strong spin-orbit coupling with a magnitude comparable to the Fermi energy, may have an odd-parity order parameter. This theoretical proposal has sparked considerable experimental studies of Cu$_x$Bi$_2$Se$_3$ and related superconductors derived from topological insulators \cite{sr-doped, sr-doped-2,  hetero}. Remarkably, two recent experimental studies on Cu$_x$Bi$_2$Se$_3$, both bulk probes of the bulk pairing symmetry, have observed in-plane uniaxial anisotropy appearing below $T_c$: Knight shift measurements \cite{zheng} and specific heat measurements in rotating field \cite{maeno2}. This provides a direct evidence of spontaneous spin rotational symmetry breaking in the superconducting state. Further theoretical analysis \cite{fu-nematic} shows that this NMR result {\it and} the absence of line nodes deduced from an earlier specific heat measurement \cite{ando} appear to be consistent only with the two-component $E_u$ order parameter, which is one of the odd-parity pairing symmetries classified in Ref.~\onlinecite{fuberg}. Very recent magnetotransport \cite{devisser} and torque magnetometry \cite{luli} measurements further support this. 

Motivated by these experimental advances, here we study the physics of odd-parity two-component superconductors. In general, superconductors with multi-dimensional order parameters may exhibit multiple superconducting phases as a function of temperature, magnetic field, pressure and chemical substitution. These phases may break time-reversal, spin rotation, or crystal symmetry, as exemplified by the $A$- and $B$-phase of $^3$He. While time-reversal-breaking (or chiral) superconductivity has been widely studied especially in the context of Sr$_2$RuO$_4$, rotational symmetry broken or \emph{nematic} superconductivity, as exhibited by Cu$_x$Bi$_2$Se$_3$, is however rare and largely unexplored. 
The main purpose of this work is to study the energetics and physical properties of two-component odd-parity superconductors, and point out the crucial role of spin-orbit coupling.

Our main results are as follows. First, we list the representative gap functions of two-component odd-parity order parameters, which differ significantly for materials with and without spin-orbit interaction. Next, by a weak-coupling analysis based on Bardeen-Cooper-Schrieffer (BCS) theory, we show that in spin-rotational-invariant materials,  the energetically favored superconducting phase below $T_c$ is rotationally invariant,  but in the presence of strong spin-orbit coupling, it can be either chiral or nematic. 
We show that the nematic phase generally has a full superconducting gap, whereas the chiral phase exhibits robust point nodes with {\it lifted} spin degeneracy on a 3D Fermi surface, resulting in low-energy Majorana-Bogoliubov quasi-particles in the bulk and topological Majorana arcs on the surface.   

{\it Two-component odd-parity order parameters---}  
Assuming that the superconducting gap is much smaller than the Fermi energy, we construct the superconducting order parameter using electron operators  on the Fermi surface: $\psi(\vec{k}) = [ c_{1}(\vec{k}),  c_{2}(\vec{k})  ]^T $. In the absence of spin-orbit coupling the index $\alpha=1,2$ labels two spin eigenstates along a global $z$ axis. In spin-orbit coupled systems, and in the presence of time-reversal ($\Theta$) and parity ($P$) symmetry, $\alpha$ is a pseudospin index labeling the two degenerate bands. 
We choose a special basis, called manifestly covariant Bloch basis (MCBB)~\cite{fu15}, in which the $[ c_1, c_2]^T$ obeys the same simple transformation properties under the symmetries of the crystal as an ordinary $SU(2)$ spinor $[c_{\uparrow}, c_\downarrow ]^T$ \cite{note1}. 

The pairing potential in BCS mean field theory of superconductivity can be explicitly expressed in MCBB as $ \hat{\Delta} =   \sum_{\vec{k}} \;  \Delta_{\alpha\beta}(\vec{k}) \epsilon_{\beta\gamma}c^\dagger_{\alpha}(\vec{k})c^\dagger_{\gamma}(-\vec{k})$, where the pairing matrix $\Delta(\vec{k})$ is basis dependent. 
Time-reversal symmetry acts as $\Theta \psi(\vec{k})  \Theta^{-1}  = i\sigma_y\psi(-\vec{k})$, and this implies for a time-reversal-invariant pairing function $(i\sigma_y)\Delta^*(\vec{k}) ( -i\sigma_y)= \Delta(-\vec{k})$.
Symmetries of the crystal point group $G$, denoted $g \in G$, act as $g \psi(\vec{k})  g^{-1} = U_g \psi(g\vec{k})$. Odd-parity pairing is then defined by the relation $\Delta(-\vec{k})=-\Delta(\vec{k})$. 

The simple transformation properties of the pairing function in the MCBB allow a straightforward classification in terms of representations of the crystal symmetry group. In spin-rotational-invariant systems, the pairing is decomposed into different representations of the symmetry group $G \times SU(2)$. For multi-component odd-parity superconducting order parameters, the pairing function is a linear combination of the basis functions, $ \Delta(\vec{k}) = \sum_{m} \vec{\xi}_{m} \Delta_{m} (\vec{k}) \cdot \vec{\sigma}$, where $m$ labels the components of the representation, and the order parameters $\vec{\xi}_{m}$ are {\it vectors} in spin space.  Because of the spin-rotational symmetry, superconducting states whose order parameters only differ by a common $SO(3)$ rotation of all vectors $\vec{\xi}_{m}$ are degenerate in energy. In contrast, in the presence of spin-orbit coupling, the electron spin and momentum transform jointly under crystal symmetry operations. Therefore, the pairing function is entirely classified by the symmetry group $G$, and $\Delta(\vec{k})$ is decomposed as $\Delta(\vec{k}) = \sum_{m} \eta_{m} \Delta_{m}( \vec{k})$ where $\eta_{m}$ are {\it scalars} and the spin structure is now fixed by $\Delta_{m}( \vec{k})$. 

\begin{table}[t]
\centering
\begin{ruledtabular}
\begin{tabular}{ccc}
Pairing form factors  & $D_{3d}$   & $D_{6h}$ \\ [4pt]
\hline  
$ \left. \begin{matrix} g_1(\vec{k}) = k_x \\ g_2(\vec{k}) = k_y \end{matrix} \; \right\} $ &  $E_{u}$& $E_{1u}$   \\[12pt]
\hline
$ \left. \begin{matrix} F^a_1(\vec{k}) = k_x\sigma_z , F^b_1(\vec{k}) = k_z\sigma_x   \\ F^a_2(\vec{k}) = k_y\sigma_z , F^b_2(\vec{k}) =k_z\sigma_y \end{matrix} \; \right\} $ &  $E_{u}$  & $E_{1u}$   \\ [12pt]
$ \left. \begin{matrix} F^c_1(\vec{k}) = k_x\sigma_y  + k_y\sigma_x  \\ F^c_2(\vec{k}) =  k_x\sigma_x - k_y\sigma_y  \end{matrix} \; \right\} $ &  $E_{u}$   &  $E_{2u}$   
\end{tabular}
\end{ruledtabular}
 \caption{Table listing the two-component odd-parity pairings to leading order $p$-wave in the harmonic expansion. We focus on hexagonal and trigonal crystal systems. The coefficients $a,b$ multiplying degenerate basis functions are arbitrary, i.e., not determined by symmetry. 
 }
\label{tab:pairing}
\end{table}

In this work we consider \emph{two}-component order parameters, i.e., $m=1,2$. Then, the pairing functions for spin-rotationally invariant and spin-orbit coupled cases take the form
\begin{align}
\Delta(\vec{k}) &= \vec{\xi}_1 g_1(\vec{k}) \cdot \vec{\sigma} + \vec{\xi}_2 g_2(\vec{k}) \cdot \vec{\sigma}, \label{eq:pairnosoc} \\
\Delta(\vec{k}) &= \eta_1 F_1(\vec{k}) + \eta_2 F_2(\vec{k}), \label{eq:pairsoc} 
\end{align}
respectively. For example, Table I shows basis functions $g_{1,2}(\vec k)$ and $F_{1,2}(\vec k)$ in the leading-order $p$-wave harmonic expansion for the trigonal $D_{3d}$ point group of Bi$_2$Se$_3$, and the hexagonal $D_{6h}$ point group of UPt$_3$. Many (but not all) of these basis functions have been obtained before, see for example Ref.~\onlinecite{yip} and references therein.   

{\it Landau theory.---}   To address the phenomenology of odd-parity pairing we consider the Ginzburg-Landau (GL) expansion of the free energy in the order parameter. We first take the spin-orbit coupled case. For the order parameters defined in Eq.~\eqref{eq:pairsoc} the free energy up to fourth order and for all symmetry groups listed in Table~\ref{tab:pairing} is given by~\cite{volovik85,sigrist91}
\begin{multline} 
F = A(T-T_c)( |\eta_1|^2 +|\eta_2|^2 )+ \\   B_{1} ( |\eta_1|^2
+|\eta_2|^2 )^2  + B_{2} | \eta_1^*\eta_2 - \eta_1\eta^*_2 |^2 \label{eq:fesoc}
\end{multline}
The GL coefficient $B_2$ decides the superconducting order below $T_c$ \cite{volovik85,sigrist91,fu-nematic}. When $B_2 < 0$ the chiral superconductor, given by $(\eta_1,\eta_2) = \eta_0 (1,\pm i)$, is favored. Chiral superconductivity, defined by nonzero $\eta_1^*\eta_2 - \eta_1\eta^*_2$, breaks time-reversal symmetry since $\Theta$ acts $\eta_i \rightarrow \eta^*_i $. When $B_2 > 0$ the nematic superconductor, given by $(\eta_1,\eta_2) = \eta_0 (\cos \theta ,\sin \theta )$, is favored. The nematic superconductor owes its name to nonzero nematic order $N_i$ given by $(N_1,N_2 ) = ( |\eta_1|^2 - |\eta_2|^2 , \eta^*_1\eta_2 + \eta^*_2\eta_1 )$. 
These components satisfy $N^*_i = N_i$ and therefore are time-reversal invariant. They transform, however, as partners of the $E_{g}$ ($D_{3d}$) and $E_{2g}$ ($D_{6h}$) representations, which implies the nematic superconductor breaks rotational symmetry. 
The nematic angle $\theta$ is pinned at three-fold degenerate discrete values only at sixth order in the GL expansion. The term $F^{(6)} = C_1 ( N^3_+ + N^3_- )$ with $N_\pm  = \eta_1 \pm i \eta_2$ discriminates the two types of nematic states with $(\eta_1, \eta_2) = \eta_0 (1,0)$ 
and $\eta_0 (0,1)$.

This should be contrasted with the Landau theory for odd-parity triplet pairing in spin-rotational invariant systems, which in terms of $(\vec{\xi}_1,\vec{\xi}_2)$ defined in Eq.~\eqref{eq:pairnosoc} is given by, at fourth order,
\begin{gather} 
F^{(4)} =   B_1(|\vec{\xi}_1|^2 + |\vec{\xi}_2|^2 )^2  + B_2|\vec{\xi}^*_1\times \vec{\xi}_1+\vec{\xi}^*_2\times \vec{\xi}_2|^2 \nonumber \\
+ B_3 |\vec{\xi}^*_1\cdot\vec{\xi}_2 - \vec{\xi}^*_2\cdot\vec{\xi}_1 |^2 + B_4 (\vec{\xi}^*_1\times\vec{\xi}_2 - \vec{\xi}^*_2\times\vec{\xi}_1 )^2 + \nonumber \\
+ B_5(N^2_1 + N^2_2) + B_6( | \vec{N}_1|^2 + | \vec{N}_2 |^2 ), \label{eq:fenosoc} 
\end{gather}
where $N_{1,2}=\xi^{a*}_{i}\tau^{z,x}_{ij}\xi^{a}_{j}$ and $N^a_{1,2} =\epsilon^{abc} \xi^{b*}_{i}\tau^{z,x}_{ij}\xi^{c}_{j}$ (repeated indices summed). The GL coefficients $B_{2}$-$B_{6}$ determine which of the four distinct superconducting states is selected immediately below $T_c$ \cite{supplmat}.


{\it Weak-coupling energetics below $T_c$.---}  To proceed, we examine the energetics of odd-parity two-component superconductors in weak-coupling BCS theory. In a microscopic theory the phenomenological GL coefficients can be evaluated as Feynman diagrams, and we exploit the symmetry of the two-component pairings to relate GL coefficients to each other and infer the relative stability of superconducting states~\cite{supplmat}.

Consider first the triplet superconductors without spin-orbit coupling. For the symmetry groups listed in Table~\ref{tab:pairing}, using the transformation properties of $(g_1,g_2)$, we find that the GL coefficients are related as $B_1=B_2=2B_5=2B_6$ and $B_3=B_4=0$. This result is obtained using only the symmetry of the form factors and holds irrespective of the Fermi surface geometry or the form of $g_{1,2}$ (i.e., order of the harmonic expansion), and leads to a very general conclusion: the rotational-invariant chiral and the helical superconductor are the favored within the weak-coupling analysis, both in $2D$ and $3D$, and they remain degenerate at fourth order in GL theory. 

\begin{figure}
\includegraphics[width=0.9\columnwidth]{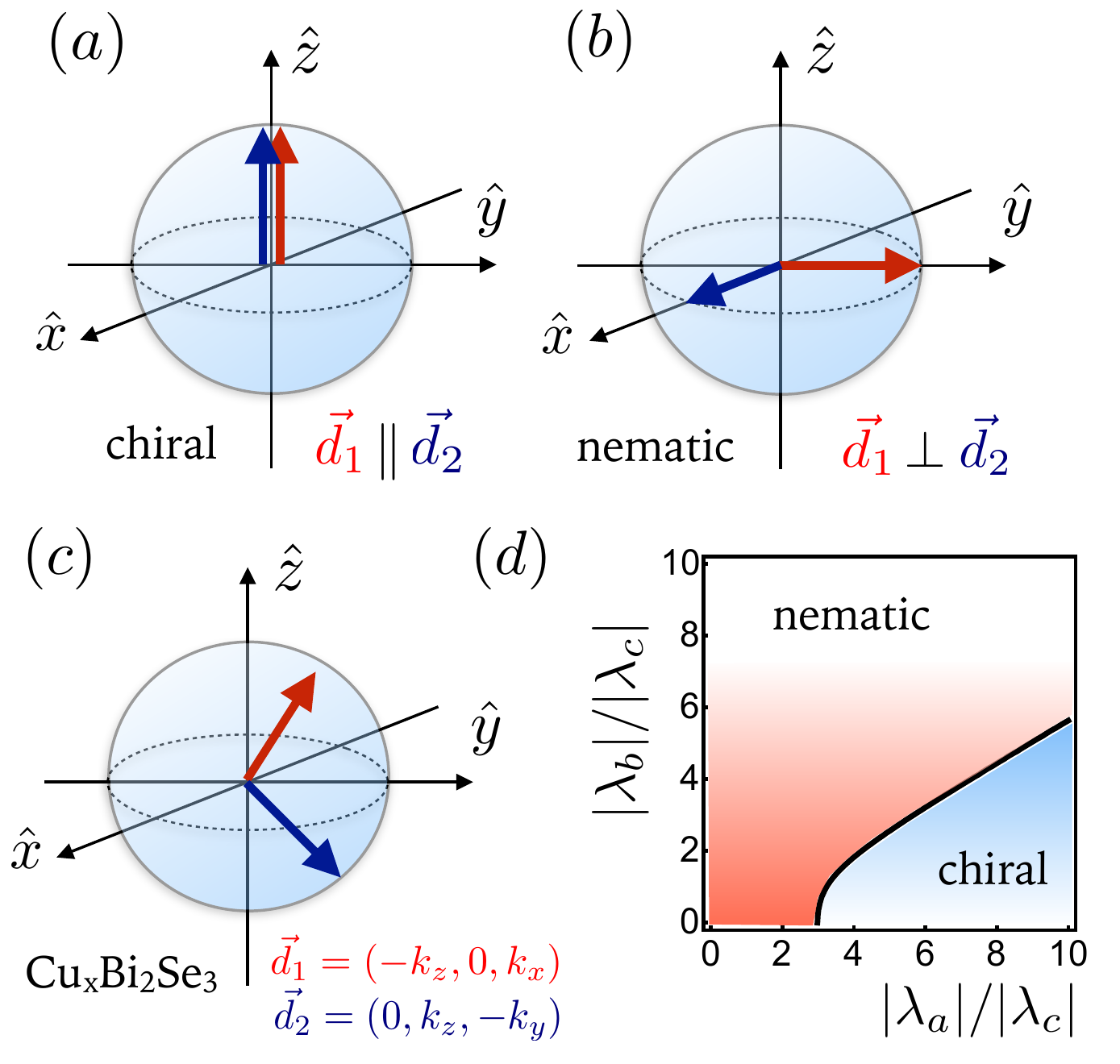}
\caption{\label{fig:dvec} (a)-(c) Pictorial representation of the geometric criterion for odd-parity two-component superconductors whose $d$-vectors $\vec{d}_{1}(\vec{k})$ and $\vec{d}_{2}(\vec{k})$ as function of $\vec k$ are (a) parallel, (b) perpendicular, and (c) have both parallel and perpendicular components. Case (c) applies to the two-orbital model of Cu$_x$Bi$_2$Se$_3$. (d) Superconducting phase diagram of odd-parity two-component superconductors with hexagonal and trigonal symmetry, obtained for pairings composed of $F^t_{1,2}(\vec k)$ with $t=1,2,3$ and assuming a spherical Fermi surface, in the $(|\lambda_a|,|\lambda_b|)/|\lambda_c|$ plane. }
\end{figure}

Next, we turn to spin-orbit superconductors described by Eq.~\eqref{eq:fesoc}, and study their energetics. 
We first derive a general expression for the GL coefficients $B_{1,2}$ and then apply the result to various gap functions given in Table~\ref{tab:pairing}. We expand the two pairing components of Eq.~\eqref{eq:pairsoc} as $F_{1,2}(\vec{k}) = \vec{d}_{1,2}(\vec k)\cdot \vec{\sigma} $ in terms of real momentum-dependent $\vec{d}_{1,2}$-vectors (which are locked to the lattice) and calculate the GL coefficients~\cite{supplmat}. Remarkably, the result for $B_2$ can be cast entirely in terms of the $d$-vector configuration on the Fermi surface, taking the simple form  
\begin{gather}
B_2 = \order{ (\vec{d}_1 \times\vec{d}_2)^2} - \order{(\vec{d}_1 \cdot \vec{d}_2)^2}, \label{eq:b2coeff}
\end{gather} 
where $ \langle ... \rangle$ is equal to an average over the Fermi surface.
Defining $I_1 = \order{(\vec{d}_1 \cdot \vec{d}_2)^2}$ and $I_2= \order{ (\vec{d}_1 \times\vec{d}_2)^2} $ we further find $B_1 = 3I_1 + I_2$. From this we obtain a general criterion for the superconducting state favored below $T_c$: the parallel component $\vec{d}_1 \parallel \vec{d}_2$ favors the chiral the superconductor whereas the orthogonal component $\vec{d}_1 \perp \vec{d}_2$ favors the nematic superconductor. A pictorial geometric representation of this is shown in Fig.~\ref{fig:dvec}(a)-(c). 

Let us now apply this result to the pairing functions of Table~\ref{tab:pairing}, and in particular to the case of Cu$_x$Bi$_2$Se$_3$. For $E_u$ pairing of $D_{3d}$ (the point group of Cu$_x$Bi$_2$Se$_3$) the gap function is a linear combination of multiple basis functions in the $p$-wave harmonic expansion, i.e., $F_{m}(\vec k) =\sum_t  \lambda_t F^t_{m} (\vec k)$ with $t=a,b,c$ and here the $\lambda_t$ are not determined by symmetry. The GL coefficients $B_{1, 2}$ will depend on the expansion coefficients $\lambda_t$ and the details of the Fermi surface~\cite{supplmat}. Assuming a three-dimensional Fermi surface, the superconducting phase diagram determined by the sign of $B_2$ is shown in Fig.~\ref{fig:dvec}(d), as a function of $|\lambda_a|/|\lambda_c|$ and $|\lambda_b|/|\lambda_c|$. The location of the material Cu$_x$Bi$_2$Se$_3$ in this phase diagram can be obtained by mapping the two-orbital model~\cite{fuberg} to the conduction band MCBB (see Suppl. Mat.). We find that $|\lambda_a | \sim |\lambda_b|$ and $\lambda_c=0$. As a result, the nematic superconductor is expected below $T_c$, consistent with the observation of rotational symmetry breaking in NMR \cite{zheng}.

In case of $E_{1u}$ and $E_{2u}$ pairing in a hexagonal crystal (i.e., symmetry group $D_{6h}$) the gap functions are similarly expanded with expansion coefficients $\lambda_t$. Now symmetry forces $\lambda_c=0$ ($E_{1u}$) and $\lambda_a=\lambda_b=0$ ($E_{2u}$), fixing the the location in phase diagram of Fig.~\ref{fig:dvec}(d). The nematic superconductor is selected as the lowest energy state independent of the Fermi surface geometry when $\lambda_a=\lambda_b=0$, since parallel component identically vanishes in this case, i.e., $\vec{d}_1 \cdot \vec{d}_2 = 0$.

The appearance of nematic superconductivity in spin-orbit coupled systems should be contrasted with triplet pairing in spin-rotationally invariant superconductors, which always leads to isotropic phases: either chiral or helical. It may also be contrasted with two-component singlet $d$-wave superconductors~\cite{nandkishore12} and spinless $p$-wave superconductors in $2D$~\cite{cheng10}: in both cases the  isotropic chiral phase ($p+ip$ and $d+id$) is favored.

{\it Gap structures.---}  We now study quasiparticle gap structures of nematic and chiral superconductors with spin-orbit coupling. First, consider the $E_u$ pairing in trigonal crystals with $D_{3d}$ point group, whose gap function takes the general form $\Delta(\vec k) =\sum_t \lambda_t (\eta_1 F^t_{1} (\vec k) + \eta_2 F^t_{2} (\vec k)) \equiv \vec d (\vec k) \cdot {\vec \sigma}$, where $F^t_{1,2}$ with $t=a,b,c$ are listed in Table~\ref{tab:pairing}.   
A nematic superconductor is obtained when $\eta_{1,2}$ and $\lambda_t$ are real.  In this case, the superconducting gap is given by $\delta (\vec k) = | {\vec d}(\vec k)|$, with $\vec k$ being on the Fermi surface. For generic values of $(\eta_1, \eta_2)$, it is vanishingly improbably to find solutions to $\vec d (\vec k)=0$, which involves {\it three} independent equations,  on the Fermi surface, which is a {\it two}-dimensional manifold. Therefore, nematic superconductors are generally nodeless \cite{fu-nematic}. Only for $\eta_2=0$, a pair of point nodes are present on the $yz$ plane, and protected by the mirror symmetry $x \rightarrow -x$ which remains unbroken in the nematic superconducting state~\cite{fu-nematic}.  

\begin{figure}
\includegraphics[width=0.9\columnwidth]{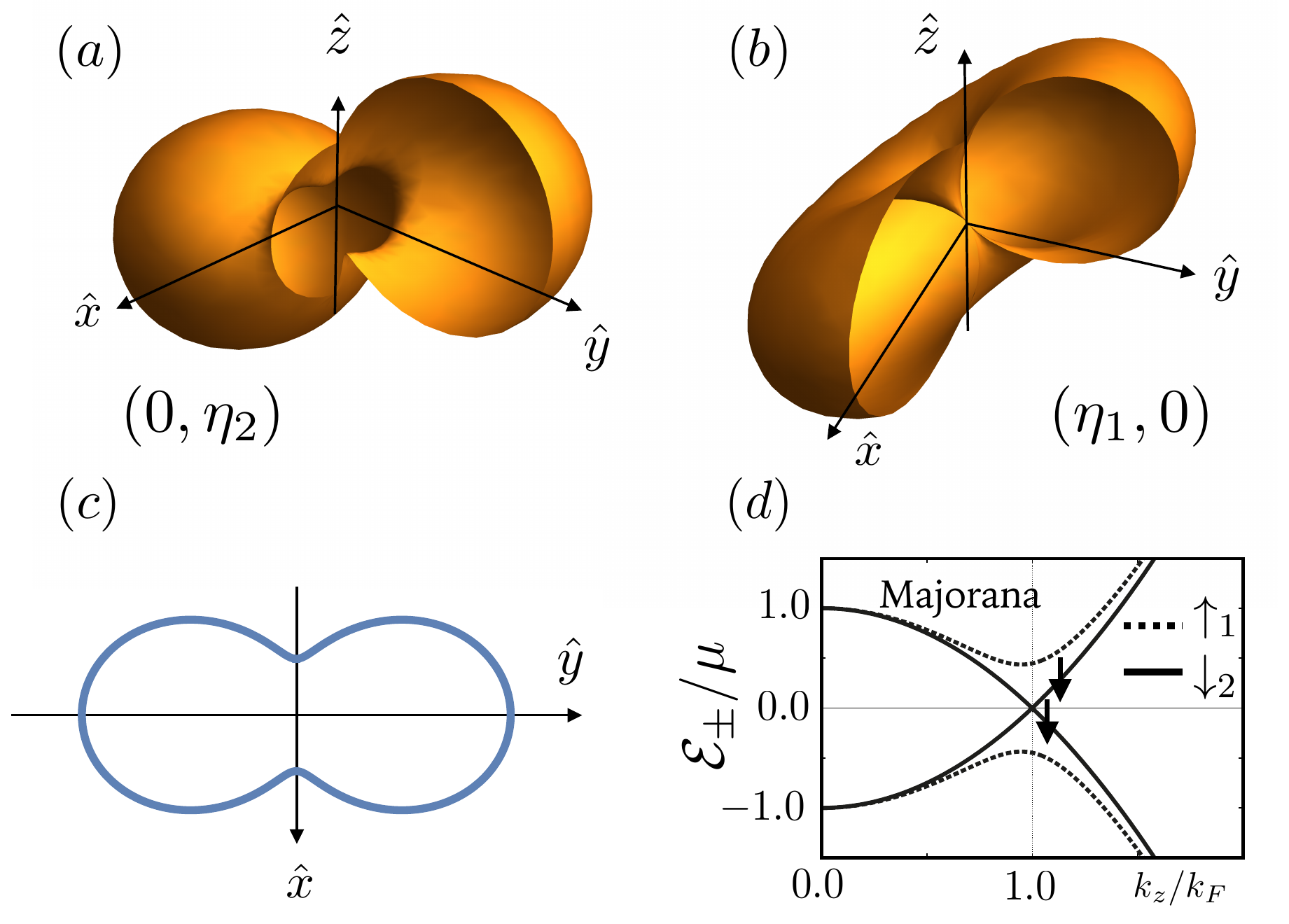}
\caption{\label{fig:gap} (a)-(b) Quasiparticle gap structures of odd-parity superconductors with $(0,1)$ and $(1,0)$ nematic components, showing the absence and presence of nodes, respectively. (c) Quasiparticle gap of $(0,1)$ nematic superconductor as function of azimuthal angle for $k_z=0$. (d) Quasiparticle spectrum $\mathcal{E}_\pm$ along $k_z$ of chiral superconductor $(1,i)$ showing Majorana node.   }
\end{figure}

The quasiparticle gap structures of these nematic superconductors are shown in Fig.~\ref{fig:gap}(a)-(b). To make a connection with Cu$_x$Bi$_2$Se$_3$ we note that experiments have reported a full pairing gap~\cite{ando}, which is consistent with a $(0,1)$ nematic state. The two-fold anisotropic gap structure of this nematic state provides a direct experimental test of the pairing symmetry of Cu$_x$Bi$_2$Se$_3$. The normal state Fermi surface has been shown to display doping dependence, becoming open and quasi-$2D$ at high doping~\cite{lahoud13}. We therefore also plot the quasiparticle gap at $k_z=0$ in Fig.~\ref{fig:gap}(c) representative for such case. The difference between open en closed Fermi surfaces may be a way to reconcile conflicting STM measurement studies~\cite{sasaki11,levy13}. 

Hexagonal crystals have higher symmetry and therefore potentially more constraints on the gap structure. In particular, certain gap function coefficients $\lambda_t$ are forced to vanish in certain pairing channels. In case of $E_{1u}$ pairing, all nematic superconductors have a pair of point nodes in the $xy$ plane due to a mirror symmetry $z \rightarrow -z$. 
In case of $E_{2u}$ pairing, the gap function $\Delta(\vec k)$ vanishes on the $z$-axis, resulting in a pair of point nodes on a 3D Fermi surface. However, for generic values of $(\eta_1,\eta_2)$, the nematic superconductor with lowered crystal symmetry allows a small admixture of a new gap function $F_{A_{1u}}(\vec k) = k_z \sigma_z$, 
whose presence leads again to a full superconducting gap. Only for special cases of $(\eta_1, \eta_2)=(\cos \theta_0, \sin \theta_0)$ with $\theta_0= (n + 1/2)\pi/3$, the presence of two mirror symmetries $x\rightarrow -x$ and $y \rightarrow -y$ protect the nodes along the $z$ axis. We have thus shown that nematic superconductors with odd-parity order parameters generally have a full gap, except for special cases associated with the presence of a mirror symmetry~\cite{supplmat}. 

In contrast, chiral superconductors with complex $\vec d$-vector $(\eta_1, \eta_2) = (1, \pm i)$ have a different gap structure. 
Of particular interest is the chiral superconductor with the $D_{3d}$ point group and $E_{u}$ pairing. From the gap function we find this pairing yields a non-unitary state with different gaps for the two pseudospin species. On a 3D Fermi surface, a particular pseudospin species determined by the chirality of the order parameter,  $\sigma_z=-1 (1)$ for the case of $\eta_2/\eta_1 = i (-i)$,  is gapless on the north and south poles $\pm \vec{K} = (0,0,\pm k_F)$, whereas the other spin species has a full gap, which is proportional to $\lambda_b$  at $\pm \vec K$. This leads to a rare case of point nodes {\it without} spin degeneracy.  As a result, the Bogoliubov quasiparticles near these two nodes form a {\it single} flavor of massless Majorana fermions in three dimensions as in particle physics. They are described by a \emph{four-component, real} quantum field $\Psi(\vec x)$, consisting of electron fields of a spin component near $\vec K$:  
\begin{gather}
\Psi^\dagger (x) = \sum_{\vec q} e^{i \vec q \cdot \vec x} \left( c^\dagger_{\vec K +\vec q}, c^\dagger_{-\vec K + \vec q},  c_{\vec K -\vec q}, c_{-\vec K -\vec q} \right).  
\end{gather}
Importantly, the field $\Psi$ lives in Nambu space and satisfies the  reality condition of Majorana fermions, $\Psi^\dagger(\vec x) =\left( \tau_x \Psi(\vec x) \right)^T$. 
The low-energy Hamiltonian for these Majorana-Bogoliubov quasiparticles, in case of $\eta_2/\eta_1=i$, is given by
\begin{gather}
H_+ = \frac{1}{2} \sum_{\vec q} \Psi^\dagger_{\vec q} \left( 
\begin{array}{cccc}
v_F q_z & 0 & 0 & v_\Delta iq_- \\
0 & -v_F q_z & v_\Delta  iq_- & 0 \\
0 & -v_\Delta  iq_+ & v_F q_z & 0 \\
-v_\Delta  iq_+ & 0 & 0 & - v_F q_z 
\end{array}
\right)  \Psi_{\vec q}, 
\end{gather}
where $q_\pm = q_x \pm i q_y$, $v_F$ is Fermi velocity in the $z$ direction, and $v_\Delta  = 2\eta_0\lambda_c$ with pairing amplitude $\eta_0$. The Hamiltonian $H_-$ for the opposite chirality $\eta_2/\eta_1= -i$ is obtained by interchanging $q_+$ and $q_-$. 
The quasiparticle dispersion near the nodes is linear in all directions, as shown in Fig.~\ref{fig:gap}(d). The presence of gapless Majorana fermions is a unique feature of chiral superconductors with spin-orbit coupling and 3D Fermi surface \cite{kozii16}.  It should be contrasted with either the nematic superconductor or $p_x \pm i p_y$ superfluid $^3$He, both of which have {\it spin-degenerate} point nodes giving rise to a four-component Dirac fermions instead of Majorana. Moreover, the chiral superconductor with $E_u$ pairing becomes fully gapped 
when the Fermi surface topology changes from a closed pocket to an open cylinder  \cite{lahoud13}.

The gap structures of both nematic and chiral superconductors, including the gap anisotropy and nodal quasiparticles, can be detected by tunneling, specific heat and thermal conductivity under a rotating field, as well as temperature- and angle-dependent London penetration depth. The chiral superconductor is topological and has chiral Majorana fermion surface states. 
When the bulk has point nodes, zero-energy surface states form a {\it single} open arc  in two-dimensional momentum space, connecting the projection of the nodes. 
Importantly, this Majorana arcs has half degrees of freedom as surface arcs in Weyl semimetals \cite{weyl} or  superconductors with spin-degenerate nodes \cite{volovik, balents, sau, zhang}. When the bulk is fully gapped, surface states consist of an array of one-dimensional chiral Majorana fermions stacked along the $z$ axis. The presence of chiral Majorana fermions on the surface gives rise to a topological thermal Hall effect, which we will study in detail elsewhere.

To summarize, odd-parity superconductivity with two-component order parameters in spin-orbit-coupled materials comes in two flavors: nematic and chiral. The relative energetics of these two phases  is determined by the spin texture of the gap function, i.e., the geometry of $d$-vectors over the Fermi surface, as shown in Eq.(\ref{eq:b2coeff}). 
The gap structures of nematic and chiral phases are obtained, and nodal quasi-particles in the latter case are identified as undoubled 3D Majorana fermions. 
Our results directly apply to a number of materials currently receiving much attention, including Cu$_x$Bi$_2$Se$_3$ and possibly Sr$_x$Bi$_2$Se$_3$\cite{sr-doped} and Nb$_x$Bi$_2$Se$_3$\cite{nb-doped}.

{\it Acknowledgment:} We thank Yoichi Ando and Yew San Hor for discussions. This work is supported by the David and Lucile Packard Foundation (LF and VK), and the Netherlands Organization for Scientific Research (NWO) through a Rubicon grant (JV).

\bibliographystyle{apsrev}

\pagebreak

\onecolumngrid

\setcounter{page}{1}

\begin{center}

{\bf \large Supplementary material for ``Odd-parity superconductors with two-component order parameters:  nematic and chiral, full gap and Majorana node''}
\newline \newline
J\"orn W. F. Venderbos, Vladyslav Kozii, and Liang Fu \\
{\it Department of Physics, Massachusetts Institute of Technology,
Cambridge, MA 02139, USA}

\end{center}

\tableofcontents

\subsection*{Overview}

This Supplementary Material (SM) is organized as follows. In Sec.~\ref{sec:oddparity} we give a more detailed introduction to odd-parity two-component pairing order parameters in different crystal systems. In particular, we present the complete expansion of pairing components in crystal harmonics, going beyond the leading order $p$-wave terms used in the main text. Then, in Sec.~\ref{sec:cubise} we give a detailed account of the application of our theory to the case of Cu$_x$Bi$_2$Se$_3$. In Sec.~\ref{sec:landau} the Landau theory of odd-parity two-component superconductors is developed in more depth. Similarities and differences with the case of tetragonal symmetry (not considered in the main text) are discussed. In Sec.~\ref{sec:bcs} we present the weak-coupling BCS approach to calculating the GL coefficients. Finally, Sec.~\ref{sec:gaps} is devoted to an extensive discussion of quasiparticle gap structures of distinct superconductors in the odd-parity two-component pairing channels. Symmetry arguments are employed to establish the protection of point node degeneracies. In addition, the low-energy theory for the nodal quasiparticles is developed and compared to previous work. 

\section{Odd-parity order parameters\label{sec:oddparity}}

In this first section we provide additional information on the definition of two-component odd-parity pairing with an emphasis on the case of spin-orbit coupling. In the main text we have focused the discussion on odd-parity two component pairing form factors in the leading order $p$-wave expansion in crystal harmonics. 
In this section of the SM we consider the general case and provide a complete list of crystal harmonics, i.e., a complete basis for the degenerate two-component channel, in which a general pairing potential can be expanded. 

In the main text we defined the electron operators in the MCBB as $\psi(\vec{k}) = [ c_{1}(\vec{k}),  c_{2}(\vec{k})  ]^T $~\cite{fu15_SM}. The normal state Hamiltonian is given by
\begin{gather}
H =  \int_{\vec{k}}  E(\vec{k})\psi^\dagger(\vec{k}) \psi(\vec{k}), \label{eq:ham}
\end{gather}
where the energy relative to the chemical potential, given by $E(\vec{k})= \varepsilon(\vec{k}) - \mu$, is a scalar function of momentum composed of terms invariant under the crystal symmetry group. It depends both on the dimensionality and crystal point group. 

The pairing potential in BCS mean-field theory can be explicitly expressed in the MCBB and takes the form
\begin{gather}
\hat{\Delta} =   \int_{\vec{k}} \;  \Delta_{\alpha\beta}(\vec{k}) \epsilon_{\beta\gamma}c^\dagger_{\alpha}(\vec{k})c^\dagger_{\gamma}(-\vec{k}), 
\end{gather}
where the pairing matrix $\Delta(\vec{k})$ is basis dependent. In the MCBB symmetries of the crystal act in a very simple way. Time reversal symmetry $\Theta$ and point group symmetries $g \in G$, where $G$ is the crystal point group, act on the electron operators $\psi(\vec{k})$ as
\begin{gather}
\Theta \psi(\vec{k})  \Theta^{-1}  = i\sigma_y\psi(-\vec{k}), \qquad g \psi(\vec{k})  g^{-1} = U_g \psi(g\vec{k}).
\end{gather}
Time reversal invariant and odd-parity superconductors ($P$ is the parity operation) then satisfy
\begin{gather}
\Theta \hat{\Delta}  \Theta^{-1} = \hat{\Delta}\;  \rightarrow \; (i\sigma_y)\Delta^*(\vec{k}) ( -i\sigma_y)= \Delta(-\vec{k}), \qquad P \hat{\Delta}  P^{-1} = - \hat{\Delta} \;  \rightarrow \; \Delta(-\vec{k})=-\Delta(\vec{k})
\end{gather}

To formulate the mean-field theory of the superconductor we define the Nambu spinor $\Phi(\vec{k}) $ in the MCBB as follows
\begin{gather} 
\Phi(\vec{k})= \begin{pmatrix} \psi(\vec{k})
 \\ i\sigma_y\psi^\dagger(-\vec{k})
\end{pmatrix} = \begin{pmatrix} c_{\alpha}(\vec{k})
 \\ \epsilon_{\alpha\beta} c^\dagger_{\beta}(-\vec{k})
\end{pmatrix}. \label{eq:nambu}
\end{gather}
In terms of the Nambu spinor the mean-field theory Bogoliubov-de-Gennes (BdG) Hamiltonian takes the form
\begin{gather}
\mathcal{H}  =  \int_{\vec{k}} \;  \Phi^\dagger(\vec{k}) \begin{pmatrix} E(\vec{k}) & \Delta(\vec{k}) \\
   \Delta^\dagger(\vec{k})   &  -E(\vec{k})  \end{pmatrix} \Phi(\vec{k}). \label{eq:bdg}
\end{gather}
Time-reversal symmetry of the normal state was assumed.

In the presence of spin-orbit coupling, when spin-rotation symmetry is broken, the superconducting pairing potential $\Delta(\vec{k})$ is decomposed into irreducible representation of the crystal symmetry group. The expansion of he pairing matrix $\Delta(\vec{k})$ in pairing channels takes the form~\cite{sigrist91_SM}
\begin{gather}
\Delta(\vec{k}) = \sum_{m} \eta_{\Gamma,m} \Delta_{\Gamma,m} (\vec{k})
\end{gather}
where $\Gamma$ labels the representation (i.e., pairing channel) and $m$ labels the components of the representation. The order parameters $\eta_{\Gamma,m}$ are complex scalars. Pairing basis functions $ \Delta_{\Gamma,m} (\vec{k})$ are obtained by decomposing products of spin and orbital angular momenta into good quantum number channels. To illustrate this, let us first consider the familiar example of full rotational symmetry $O(3)$, i.e, the absence of crystal anisotropy, in which case total angular momentum represents the good quantum number. The addition of an $L=1$ orbital angular momentum $\vec{k}$, and an $S=1$ spin angular momentum $\vec{\sigma}$ gives total angular momentum $J=0,1,2$, which label the pairing channels (i.e., $\Gamma_{J=0,1,2}$). Specifically, the spin-orbit coupled total angular momenta take the form
\begin{align}
J=0 \qquad &\mathcal{O} = k_i\sigma_j\delta_{ij},  \nonumber \\
J=1 \qquad & \mathcal{P}_i = k_j\sigma_k\epsilon_{ijk},  \nonumber \\
J=2 \qquad &\mathcal{Q}_{ij} = k_i\sigma_j - 2\vec{k}\cdot \vec{\sigma}/3,
\end{align}
representing a scalar, a vector, and a rank-2 tensor. 
%

In a crystal solid, the full rotational symmetry $O(3)$ is reduced to the crystal symmetry group $G$ and the electron spin is locked to the lattice. The pairing potential $\Delta(\vec k)$ is decomposed into pairing basis functions $ \Delta_{\Gamma,m} (\vec{k})$ labeled by representations of the crystal point group. Instead of total angular momentum, representations of the crystal group are the pairing channels. Clearly, the form of the pairing functions $ \Delta_{\Gamma,m} (\vec{k})$ then depends on the crystal system. In this work we study two-component superconductors, and we are therefore interested in crystal systems that admit odd-parity two-component pairing channels, i.e., odd-parity twofold degenerate representations. In order to compare with the case of vanishing spin-orbit coupling, we focus on crystal systems which also admit a two-component odd-parity triplet pairing channel (with spin-rotation symmetry, $G \times SU(2)$). Common crystal point groups satisfying these conditions are $G=D_{6h}$ (hexagonal), $G=D_{4h}$ (hexagonal) and $G=D_{3d}$ (trigonal). In the main text we exclusively focus on the hexagonal and trigonal crystal systems, since these cases can be described by the same theory. Moreover, the material class we have in mind as a direct application, exemplified by the case of Cu$_x$Bi$_2$Se$_3$, has trigonal symmetry. Differences arise in case of tetragonal symmetry, which we will discuss in this SM, specifically in Secs.~\ref{sec:landau} and~\ref{sec:bcs}. In a nutshell, the difference arises from the fact that \emph{products} of two-component representations in the tetragonal system do not contain two-component irreducible representations. 

\begin{table}[t]
\centering
\begin{tabularx}{0.35\textwidth}{XXXXX}
\hline \hline
   & & $D_{3d}$  & $D_{4h}$ &  $D_{6h}$ \\ [4pt]
\hline  
$(k_x,k_y )$ &  &  $E_{u}$&    $E_{u}$  &$E_{1u}$ \\ [2pt]
$k_z$   & & $A_{2u}$ & $A_{2u}$ &  $A_{2u}$ \\ [2pt]
$(\sigma_x,\sigma_y )$  & & $E_{g}$ & $E_{g}$ & $E_{1g}$\\[2pt]
$\sigma_z$ & &  $A_{2g}$ &  $A_{2g}$ &  $A_{2g}$ \\
\hline\hline
\end{tabularx}
 \caption{Table listing the symmetry of leading order $L=1$ ($p$-wave) and $S=1$ spin components in the hexagonal, tetragonal and trigonal crystal systems. }
\label{tab:irreps}
\end{table}


In case of spin-orbit coupled materials, leading order spin-orbit coupled crystal lattice harmonics are obtained by taking products of $p$-wave orbital angular momenta and spin. The symmetry labels of the (leading order) $L=1$ ($p$-wave) momenta $k_i$ and spin Pauli matrices $\sigma_i$ in the crystal symmetry groups are summarized in~\ref{tab:irreps}. This yields the pairing functions listed in Table I of the main text. Here we illustrate this based on the example of the crystal group $D_{3d}$. Taking the product of $(k_x,k_y )$ and $(\sigma_x,\sigma_y )$, $(k_x,k_y )$ and $\sigma_z$, as well as $k_z$ and $(\sigma_x,\sigma_y )$, we have, respectively,
\begin{align}
 & E_u \times E_g = A_{1u} + A_{2u} + E_{u} ,\\
&  E_u \times A_{2g} = E_{u}  , \\
&  A_{2u} \times E_{g} = E_{u}.
\end{align}
The right hand side of the products of representations, which is the lattice analog of total angular momentum, contains the two-component odd-parity pairings with $E_u$ symmetry. For instance, the $E_u$ term of first line corresponds to the pairing $(k_x\sigma_x - k_y\sigma_y , k_x\sigma_y  + k_y\sigma_x )$. The second and third lines correspond to the pairings $ ( k_x, k_y)\sigma_z$ and $k_z ( \sigma_x,\sigma_y)$, respectively. 

Note that all these pairings have the same symmetry, namely $E_u$ symmetry. This is a manifestation of the fact the $E_u$ channel is highly degenerate, already at the level of the leading order $p$-wave expansion. In general, for any odd-parity two-component pairing channel, such as $E_u$ in case of point group $D_{3d}$, arbitrary linear combinations of all admissible pairing basis functions should be considered. The most general pairing matrix in a given channel is an expansion in degenerate basis functions to arbitrary order, and the expansions coefficients are not determined by symmetry. Even though the degeneracy is infinite, there exists a notion of completeness. More specifically, given a degenerate representation $\Gamma$, the pairing component $\Delta_{m} (\vec{k})$ of such representation can be expanded in a complete and finite set of harmonics $ \Delta_{l,m} (\vec{k})$ as~\cite{yip93_SM}
\begin{gather}
\Delta_{m} (\vec{k}) = \sum_{l} \lambda_{l} \mathcal{I}_l (\vec{k}) \Delta_{l,m} (\vec{k}),  \label{eq:expand}
\end{gather}
where $\mathcal{I}_l (\vec{k})$ is any function fully invariant under the crystal symmetry group. The set $ \Delta_{l,m} (\vec{k})$ is finite, but $\Delta_{m} (\vec{k}) $ is still infinitely degenerate due to the functions $\mathcal{I}_l (\vec{k})$. For each crystal system and representation we can list the complete set of functions $\Delta_{l,m} (\vec{k})$. 

Starting with the group $D_{6h}$, the degenerate set of two-component basis functions of $E_{1u}$ symmetry are given by the real and imaginary parts of the followings functions
\begin{gather*}
E_{1u} \; : \quad k_+ \hat{z}, \; k_z \hat{r}_+, \; k_z k^2_+\hat{r}_-, \; k^5_- \hat{z},\; k_z k^4_-\hat{r}_-,\;  k_z k^6_-\hat{r}_+,
\end{gather*}
where $k_\pm = k_x \pm i k_y$ and $\hat{r}_\pm = \hat{x} \pm i \hat{y}$. The degenerate set of basis functions of $E_{2u}$ symmetry is given by the real and imaginary parts of the harmonics
\begin{gather*}
E_{2u} \; : \quad  k_+ \hat{r}_+, \; k^2_+k_z \hat{z}, \; k^4_-k_z \hat{z}, \; k^3_+ \hat{r}_-, \; k^3_- \hat{r}_-,\; k^5_-\hat{r}_+.
\end{gather*}
In case of crystal symmetry $D_{3d}$ these two channels merge into a single channel with $E_u$ symmetry, and linear combinations of both sets are allowed. This degeneracy, which follows from lower symmetry, is an important property of the crystal system $D_{3d}$. The complete list of degenerate functions is given by the following functions
\begin{gather*}
E_{u} \; : \quad  k_+ \hat{z}, \; k_z \hat{r}_+, \; k_z k^2_+\hat{r}_-, \; (k^3_+ + k^3_-)\hat{r}_+, \; i k_- \hat{r}_-, \; ik^2_-k_z \hat{z}.
\end{gather*}
Again, the two degenerate partners of the representation are obtained by taking the real and imaginary parts. It is important to point out that one has to take $k_- \hat{r}_-$ instead of $k_+ \hat{r}_+$ in order to have the correct angular momentum and transform correctly under rotations. In addition, the extra factor $i$ makes sure that partners are switched so as to transform correctly under the non-principle two-fold rotations. The same is true for $ik^2_-k_z \hat{z}$, which also has $E_{2u}$ symmetry in the hexagonal system.

For completeness of presentation, in case of tetragonal symmetry $D_{4h}$ the set of basis functions of the odd-parity two-component representation $E_u$ are given by
\begin{gather*}
E_{u}\; : \quad  \begin{pmatrix} k_x  \\ k_y \end{pmatrix}\hat{z}, \begin{pmatrix} k^3_x  \\ k^3_y \end{pmatrix}\hat{z}, k_z\begin{pmatrix} \hat{x}   \\ \hat{y} \end{pmatrix},k_z\begin{pmatrix} k^2_x\hat{x}   \\ k^2_y\hat{y} \end{pmatrix}, 
k_xk_yk_z\begin{pmatrix} \hat{y}   \\ \hat{x} \end{pmatrix},k_xk_yk_z\begin{pmatrix} k^2_x\hat{y}   \\ k^2_y\hat{x} \end{pmatrix}.
\end{gather*}


To conclude this section, we consider the case of no spin-orbit coupling, so as to compare the two cases in various subsequent sections of this SM. When spin-orbit coupling is vanishingly weak, the pairing is decomposed with respect to the symmetry group $G \times SU(2)$. For odd-parity pairing the pairing matrix is written as
\begin{gather}
\Delta(\vec{k}) = \sum_{m} \vec{\xi}_{\Gamma,m} \Delta_{\Gamma,m} (\vec{k}) \cdot \vec{\sigma} =  \vec{d}(\vec{k}) \cdot \vec{\sigma}, \label{eq:pairnosoc}
\end{gather}
and, importantly, the $d$-vector, given by $\vec{d}(\vec{k}) =  \sum_{m} \vec{\xi}_{\Gamma,m} \Delta_{\Gamma,m} (\vec{k}) $, is free to rotate in spin space as a result of $SU(2)$ symmetry, whereas in case of spin-orbit coupling it is locked to the lattice.

\section{Application to Cu$_x$Bi$_2$Se$_3$: mapping to MCBB\label{sec:cubise}}

In the main text we consider the general case of odd-parity two-component superconductors using the Fermi surface pseudospin MCBB. As we explain in the main text, a specific and highlight example of a material for which odd-parity pairing is relevant is Cu$_x$Bi$_2$Se$_3$ (CuBiSe). CuBiSe is a doped topological insulator material and therefore generally described by a generic Hamiltonian explicitly taking orbital and spin degrees of freedom into account~\cite{fuberg_SM}. Here we demonstrate how such description can be mapped to a two-band model in the MCBB basis.

The two-\emph{orbital} model is expressed in the basis $c(\vec{k}) = [ c_{a\up}(\vec{k}),  c_{a\down}(\vec{k}) , c_{b\up}(\vec{k}),  c_{b\down}(\vec{k})  ]^T $, where $a,b$ label the orbitals. The general low-energy $\vec{k}\cdot \vec{p}$ Dirac Hamiltonian $H_0(\vec{k})$ is given by 
\begin{gather}
H_0(\vec{k}) =  v(  k_y\sigma_x\tau_z-k_x\sigma_y\tau_z) +v_z k_z \tau_y+m\tau_x,
\end{gather}
where the $\sigma_i$ are Pauli matrices acting on spin $\up,\down$ and $\tau_i$ are Pauli matrices acting on the orbital degree of freedom ($\tau_z=\pm 1$, with $\pm 1= a,b$).

The low-energy Dirac Hamiltonian has an artificial full rotational invariance to linear order in $\vec{k}$. Terms of higher order momentum will reduce that symmetry to the crystal symmetry group $D_{3d}$, which is the true symmetry group of the material~\cite{fu-nematic_SM}. A third order term fully invariant under crystal symmetry, which generates hexagonal warping of the Fermi surface, is given by
\begin{gather}
H'_0(\vec{k}) = -\lambda(k_+^3+k_-^3) \sigma_z\tau_z.
\end{gather}
In order to obtain the correct gap structures in the two-orbital model it is crucial to include this hexagonal warping term. 

In the two-orbital model the Nambu spinor takes the form
\begin{gather}
\Phi(\vec{k}) = \begin{pmatrix} c_{i \alpha}(\vec{k})
 \\ \epsilon_{\alpha\beta}c^\dagger_{i\beta}(-\vec{k})
\end{pmatrix}, \label{eq:nambu2}
\end{gather}
where $i$ runs over the orbital degrees of freedom and $\alpha$ over spin. In this basis the BdG mean-field Hamiltonian reads
\begin{gather}
\mathcal{H}(\vec{k})\equiv  \begin{pmatrix} H(\vec{k})
  & \Delta \\
\Delta^\dagger & -H(\vec{k})\end{pmatrix}, \label{eq:orbitalbdg}
\end{gather}
where, again, time-reversal symmetry $\Theta$ was used.

\subsection{Classification of pairing}

In order to classify the superconducting channels by crystal lattice symmetry we review the action of point symmetries.  Let $R \in D_{3d}$ be a point group symmetry of $H_0$. Then one has the relation $U_R  H_0(\vec{k}) U^\dagger_R = H_0(R\vec{k})$. Note that in principle $U_R $ may depend on $\vec{k}$, but we do not explicitly need this. In the Nambu basis \eqref{eq:nambu2} the BdG Hamiltonian transforms as 
\begin{gather}
 \begin{pmatrix}  U_R &  \\
   &U_R    \end{pmatrix}   \begin{pmatrix} H_0(\vec{k}) &  \Delta \\
 \Delta^\dagger  & -H_0(\vec{k})  \end{pmatrix}  \begin{pmatrix}   U^\dagger_R &  \\
  &   U^\dagger_R \end{pmatrix} 
= \begin{pmatrix} H_0(R\vec{k}) & U_R \Delta U^\dagger_R \\
U_R \Delta^\dagger U^\dagger_R   & -H_0(R\vec{k})  \end{pmatrix}.
\end{gather}
If the pairing potential is symmetric under the $R$ we have $U_R \Delta U^\dagger_R =\Delta $. Note that we do not consider momentum-dependent pairings in the orbital basis. 

The symmetry group $D_{3d}$ is generated by three elements, a threefold rotation $C_3$, a twofold rotation $C'_2$ and
parity $P$. Explicitly, the matrices of these generators are given by 
\begin{align}
P &\rightarrow U_{P}  = \tau_x, \nonumber \\
C_3 & \rightarrow U_{C_3}  = e^{-i\pi\sigma_z/3}, \nonumber \\
C'_2 & \rightarrow U_{C'_2}  = -i\sigma_x\tau_x. 
\end{align} 
Note that this implies a mirror reflection symmetry $M_{yz}$ in the $y-z$ plane given by $M_{yz} = PC'_2 =  -i\sigma_x$. 

\begin{figure}
\includegraphics[width=0.45\textwidth]{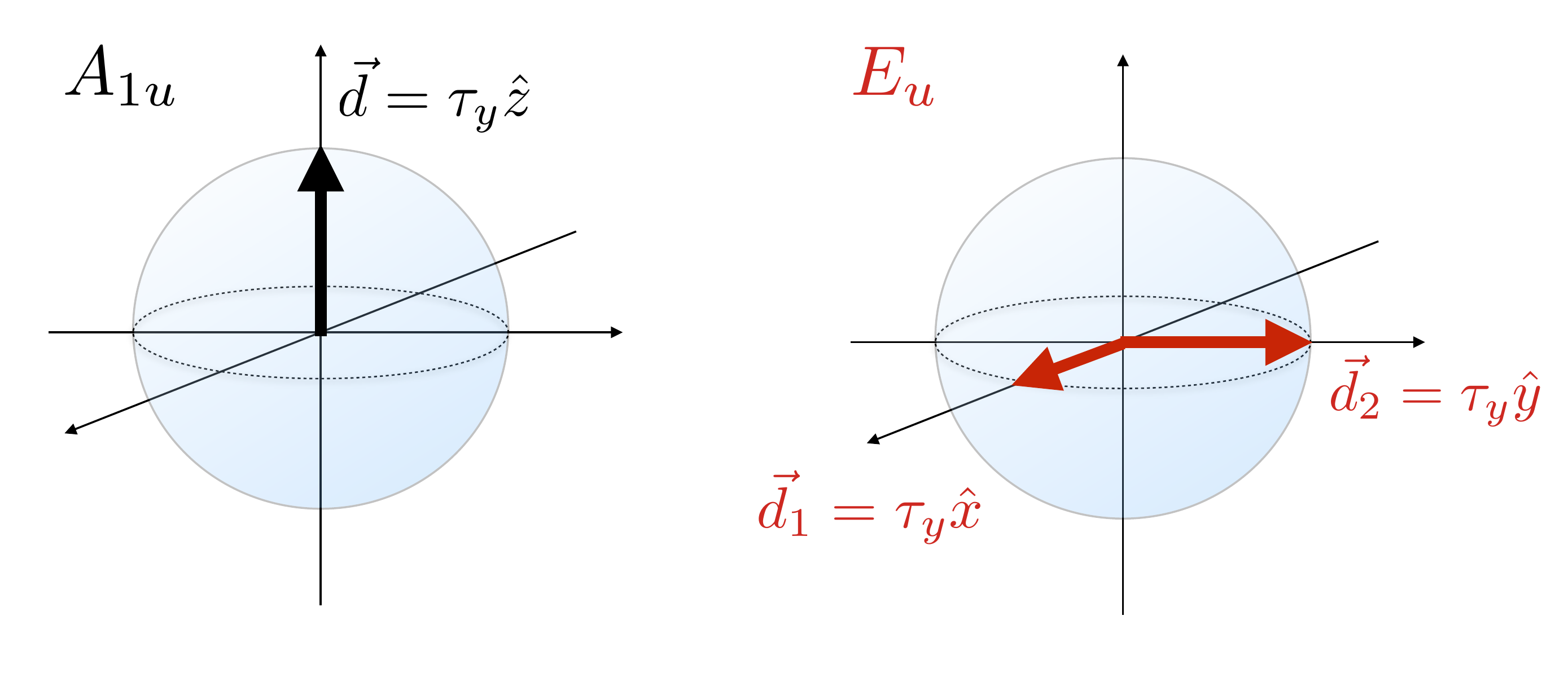}
\caption{\label{fig:pairingcubise} Graphical and geometric representation of the odd-parity triplet pairings with $A_{1u}$ and $E_u$ symmetry. Since triplet pairings can be written as $\Delta = \vec{d} \cdot \vec{\sigma} (i\sigma_y)$, they can be represented by their $\vec{d}$-vectors. The pairings with $A_{1u}$ and $E_u$ symmetry correspond to inter-orbital ($\sim \tau_y$) easy-axis (i.e., $z$) and easy-plane (i.e., $x-y$) pairing, respectively. }
\end{figure}

With these symmetry transformation properties of the pairing matrices all possible pairings given by $\tau_i\sigma_j$ can be classified according to point group symmetry. One finds three odd-parity pairing channels, with $A_{2u}$, $A_{1u}$ and $E_u$ symmetry. The $E_u$ channel is twofold degenerate and therefore constitutes a two-component pairing representation. The pairing matrices corresponding to these channels are listed in Table~\ref{tab:mcbbmapping}. These can be used to write down explicit expressions for the pairing operators. The $A_{2u}$ pairing operator takes the form
\begin{gather}
A_{2u} :\quad  c^\dagger_{i\alpha} (\tau_z)_{ij} (i \sigma_y)_{\alpha\beta} c^\dagger_{j\beta} = c^\dagger_{a\up} c^\dagger_{a\down}
-c^\dagger_{b\up} c^\dagger_{b\down}.
\end{gather}
This is a spin-singlet pairing state with a different sign for the two orbitals. Instead, $A_{1u}$ pairing is spin-triplet and inter-orbital, given by
\begin{gather}
A_{1u} :\quad  c^\dagger_{i\alpha}  (\tau_y)_{ij}(i \sigma_z\sigma_y)_{\alpha\beta}c^\dagger_{j\beta} = c^\dagger_{a\up} c^\dagger_{b\down}
+c^\dagger_{a\down} c^\dagger_{b\up}.
\end{gather}
The two-component $E_u$ pairing is inter-orbital and spin-triplet, but in contrast to $A_{1u}$, which is polarized along $z$, the two components are in the $x-y$ plane. Explicitly, $E_u$ pairing is given by
\begin{gather}
E_u :\quad \left\{  \begin{matrix} c^\dagger_{i\alpha}  (\tau_y)_{ij}(i \sigma_x\sigma_y)_{\alpha\beta} c^\dagger_{j\beta} = i (c^\dagger_{a\up} c^\dagger_{b\up} -c^\dagger_{a\down} c^\dagger_{b\down} ) \\
c^\dagger_{i\alpha}  (\tau_y)_{ij}(i \sigma_y\sigma_y)_{\alpha\beta} c^\dagger_{j\beta} = c^\dagger_{a\up} c^\dagger_{b\up}
+c^\dagger_{a\down} c^\dagger_{b\down}
 \end{matrix} \right.
\end{gather}
Schematically, the odd-parity spin-triplet pairings are shown in Fig.~\ref{fig:pairingcubise} in terms of orbital-matrix dependent $d$-vectors. 

The full pairing operator $\Delta$ of Eq.~\eqref{eq:orbitalbdg} is written in terms of the complex order parameters $\eta_{1,2}$ as
\begin{gather}
 \Delta =  \Delta_1 + \Delta_2= \eta_1 \tau_y\sigma_x + \eta_2 \tau_y\sigma_y .
\end{gather}

\subsection{Mapping to the MCBB of the conduction band}

The next step is to perform the mapping from the two-orbital (four-band including spin) model to an effective two-band model. The thrust of the mapping is to project all operators onto the conduction band and ignore the valence band. We start by expanding the electron operators $c(\vec{k}) $ in band eigenoperators $\psi_{\alpha}^i(\vec{k})$, where $i=c,v$ labels the conduction and valence band, and $\alpha=1,2$ are the band indices which will become the pseudospin indices in the MCBB~\cite{kozii_SM,Yip_SM}. We write
\beq
c(\vec{k}) = \sum_{\alpha i} a_{\alpha}^i(\vec{k}) \psi_{\alpha}^i(\vec{k}). \label{Sc}
\eeq
Since the energy bands are doubly degenerate, the choice of band eigenstates is not unique. We adopt a specific basis, the MCBB mentioned in the main text and recently introduced in~\cite{fu15_SM}, where the band eigenstates are chosen to be fully spin polarized along the $z$ direction at the origin of point group symmetry operations. In the two-orbital model considered here, the bonding orbital, which is the eigenstate of $\tau_x$ operator with eigenvalue $+1$, is invariant under all symmetry operations, thus playing the role of the coordinate origin. With this we obtain the eigenvectors in the MCBB, $a_{ \alpha}^c(\vec{k})$ corresponding to the states of the conduction band, which have the form
\beq
a_{ 1}^c(\vec{k})=\left( \begin{array}{c} \beta_+ - i \hat k_z \beta_-\\ i\hat k_+ \beta_- \\ \beta_+ + i \hat k_z \beta_-\\ -i\hat k_+ \beta_- \end{array}   \right), \quad a_{2}^c(\vec{k})=\left( \begin{array}{c}  -i \hat k_-\beta_- \\ \beta_+ - i \hat k_z \beta_- \\ i\hat k_-\beta_-\\ \beta_+ + i \hat k_z \beta_-   \end{array}   \right), \label{SMCBB}
\eeq
where we have used the definitions
\beq
\hat k_{\pm} = \hat k_x\pm i \hat k_y, \quad \beta_{\pm}=(1/2)\sqrt{1\pm (m/\mu)}, \quad \mu=\sqrt{m^2+\tilde{v}^2 \kf^2}, \quad \tilde{v}^2 \kf^2 \equiv v^2 (k_x^2 + k_y^2) + v_z^2 k_z^2
\eeq
Mathematically, the mapping onto the conducting band simply proceeds by keeping $a_{\alpha}^c(\vec{k})$ in Eq.~\eqref{Sc} only, and omitting the terms coming from the conduction band $a_{\alpha}^v(\vec{k})$. As an example, for the non-degenerate $A_{1u}$ pairing we obtain
\beq
c^\dagger_{i\alpha}  (\tau_y)_{ij}(i \sigma_z\sigma_y)_{\alpha\beta}c^\dagger_{j\beta} \quad \rightarrow \quad  \tilde F_1(\hat{k}) = \sum_j \psi^\dagger_{\alpha} (i \hat{k}_j\sigma_j\sigma_y)_{\alpha\beta}\psi^\dagger_{\beta}
\eeq
For all other channels the mapping to the conduction band MCBB is listed in Table~\ref{tab:mcbbmapping}. In particular, for $E_u$ pairing we find that 
\beq
F_2^{x}(\hat{k})= \hat k_x \sigma_z - \hat k_z \sigma_x, \quad F_2^{y}(\hat{k})= \hat k_y \sigma_z - \hat k_z \sigma_y,
\eeq
which is a specific linear combination of the p-wave harmonic basis functions $k_i\sigma_z$ and $k_z\sigma_i$ denoted $F^a_{1,2}$ and $F^b_{1,2}$ in the main text.

\begin{table}[t]
\centering
\begin{tabularx}{0.8\textwidth}{XXX}
    \hline  \hline 
 Symmetry  & Full orbital  & Band MCBB  \\ [4pt] \hline
$ A_{1u}$ &   $ \tau_y \sigma_z $ & $\tilde F_1(\hat{k})  =  \hat k_x \sigma_x + \hat k_y \sigma_y + \hat k_z \sigma_z $\\ [4pt]\hline
 $ E_u $  & $ \tau_y \sigma_x $ & $\tilde F_2^{x}(\hat{k})= \hat k_x \sigma_z - \hat k_z \sigma_x$\\[4pt] 
   &$ \tau_y \sigma_y$  & $\tilde F_2^{y}(\hat{k})= \hat k_y \sigma_z - \hat k_z \sigma_y$ \\[4pt]
 $A_{2u}$ & $\tau_z$  & $\tilde F_2^{z}(\hat{k})=\hat k_x \sigma_y -\hat k_y \sigma_x $\\  [4pt]
       \hline \hline
\end{tabularx}
\caption{Table showing the symmetry classification of pairings in CuBiSe, both in the full orbital basis (in terms of orbital matrices $\tau_i$ and spin $\sigma$) and in the two-band MCBB. We note that in the full orbital basis, $\sigma$ acts on the electron spin $\up,\down$ and in the MCBB $\sigma$ acts on the pseudospin $1,2$. Here we have defined $\hat k_{x(y)}=vk_{x(y)}/\sqrt{v^2(k_x^2+k_y^2)+v_z^2k_z^2}$ and  $\hat k_{z}=v_zk_z/\sqrt{v^2(k_x^2+k_y^2)+v_z^2k_z^2}$.}
\label{tab:mcbbmapping}
\end{table}

\section{Landau theory of two-component odd-parity superconductors\label{sec:landau}}

\subsection{General Ginzburg-Landau free energy}

The GL free energy density is an expansion in the order parameter fields to given order and depends on representation of the order parameters in the crystal symmetry group. For spin-orbit coupled odd-parity two-component superconductors of $E_u$ symmetry (point group $D_{3d}$), as well as $E_{1u}$ and $E_{2u}$ symmetry (point group $D_{6h}$), the GL free energy density expanded up to sixth order in the order parameters $(\eta_1, \eta_2)$ is given by
\begin{gather}
F = A(T-T_c)( |\eta_1|^2 +|\eta_2|^2 )+B_{1} ( |\eta_1|^2 +|\eta_2|^2 )^2   + B_{2} | \eta_1^*\eta_2 - \eta_1\eta^*_2 |^2 + C_1(N^3_+ + N^3_-) +C_{2} ( |\eta_1|^2 +|\eta_2|^2 )^3 \nonumber \\
+ C_{3} ( |\eta_1|^2 +|\eta_2|^2 )| \eta_1^*\eta_2 - \eta_1\eta^*_2 |^2. \label{eq:fehexatrig}
\end{gather}
This free energy is studied in the main text, the additional sixth order terms with the GL coefficients $C_{2,3}$ have no qualitative impact on the physics. We mention here that the number of independent invariants depends on the crystal system. In this SM (see below) we comment on the free energy describing spin-orbit coupled odd-parity two-component superconductivity in a tetragonal material, which differs from~\eqref{eq:fehexatrig}. 

\begin{figure}
\includegraphics[width=0.5\textwidth]{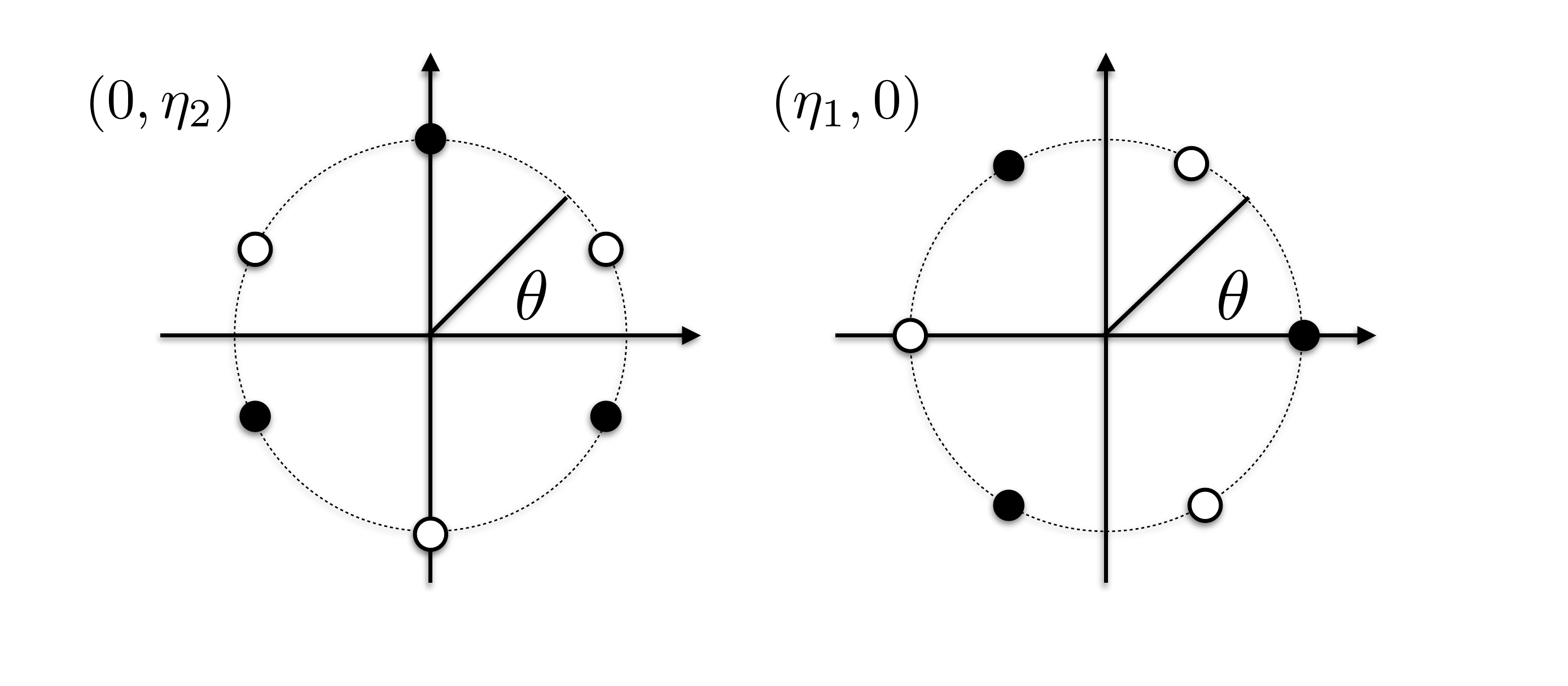}
\caption{\label{fig:nematicsolutions} Figure of nematic superconducting solutions $(\eta_1,\eta_2) = \eta_0(\cos\theta,\sin\theta)$ of~\eqref{eq:fehexatrig}. (Left) Solutions when $C_1>0$ given by $\theta-\pi/2=0,2\pi/3,4\pi/3 $, an example of which is $(\eta_1=0,\eta_2)$. (Right) Solutions when $C_1>0$ given by $\theta =0,2\pi/3,4\pi/3 $, an example of which is $(\eta_1,\eta_2=0)$. Note that solutions which differ by an overall sign are identified (black and white dots).}
\end{figure}

The subsidiary order parameters of the superconducting states are obtained by taking the second order product of representations as
\begin{gather} 
E^*_u \times E_u = A_{1g} + A_{2g} + E_g.
\end{gather} 
Here $E_u$ and $E_g$ apply to trigonal symmetry, but a similar result is obtained for hexagonal symmetry. Order parameter combinations transforming as $A_{2g}$ and $E_g$ are given by
\begin{gather} 
A_{2g}  \;   \rightarrow \; - i (\eta_1^*\eta_2 - \eta_1\eta^*_2) =  \frac{1}{2}(| \eta_+|^2  - | \eta_-|^2 ), \qquad E_g \; \rightarrow \; \left\{  \begin{matrix}  | \eta_1|^2  - | \eta_2|^2  \\
- (\eta_1^*\eta_2 + \eta_1\eta^*_2)\end{matrix}. \right. \label{eq:subs}
\end{gather} 
The nematic $E_g$ components are used to define $N_+ = N_1 +i N_2 = \eta^*_+\eta_- $, where $\eta_+ = \eta_1 + i \eta_2$. The sixth order invariant $N^3_+ + N^3_-$ therefore represents $(\eta^*_+\eta_- )^3 + (\eta^*_-\eta_+ )^3$. To see how this term discriminates different nematic superconducting solutions we write the general nematic solution as $\eta_+ = \eta_0 e^{i\theta}$. Then, $C_1(N^3_+ + N^3_-)$ is given by $ C_1 \eta^6_0 \cos 6\theta$. Depending on the sign of $C_1$ the solutions $\cos 6\theta = 1$ ($C_1 < 0$) or $\cos 6\theta = -1$ ($C_1 > 0$) have lower energy. The continuous degeneracy of $\theta$ at fourth order is therefore lifted and a discrete degeneracy results, where $\theta = \pi n/3$ or $\theta = \pi/2 + \pi n/3$. This is shown schematically in Fig.~\ref{fig:nematicsolutions}.

\subsubsection{Coupling to other representations}

In addition to a coupling to magnetic and nematic order, the odd-parity two-component superconducting order parameters can couple to superconductors with different symmetry. Clearly, there will be a fourth order coupling in even powers of the order parameters of the two distinct representations of the general form $|\chi|^2 |\eta|^2$, where $\chi$ is another superconducting order parameter. Here we are interested in coupling terms consisting of odd powers of the respective order parameters. We first distinguish the $E_{1u}$ and $E_{2u}$ channels of the hexagonal symmetry group. In case of $E_{2u}$ symmetry, the couplings of interest are given by the invariant terms in the decomposition of $A_{1u}\times E^*_{2u}\times E_{2u}\times E^*_{2u} $ and $A_{2u}\times E^*_{2u}\times E_{2u}\times E^*_{2u} $. Instead, in case of $E_{1u}$ symmetry the couplings follow from $B_{1u}\times E^*_{1u}\times E_{1u}\times E^*_{1u} $ and $B_{2u}\times E^*_{1u}\times E_{1u}\times E^*_{1u} $.

Let us take the $E_{2u}$ channel as an example. We denote the order parameter corresponding to $A_{1u}$ symmetry as $\chi$, and the order parameter with $A_{2u}$ symmetry as $\chi'$. Then the coupling of $A_{1u}$ and $E_{2u}$ gives rise to a contribution to the free energy $F[\chi,\eta_i] $ which takes the form
\begin{gather}
F[\chi,\eta_i] = e^{i\gamma} \chi [\eta^*_1 (|\eta_1|^2-|\eta_2|^2) - \eta^*_2 (\eta_1\eta^*_2+\eta_2\eta^*_1 ) ] + \text{c.c}  =  e^{i\gamma} \chi (\eta^*_1 N_1 + \eta^*_2 N_2 ) + \text{c.c} , \label{eq:glmixing}
\end{gather}
where $\gamma$ is an arbitrary phase factor. As before, $N_{1,2}$ are the nematic components derived from the $\eta_{1,2}$ order parameters given in Eq.~\eqref{eq:subs}. Similarly, the coupling of $\chi'$ and $\eta_{1,2}$ is given by
\begin{gather}
F[\chi',\eta_i] =e^{i\gamma'} \chi' [- \eta^*_1 (\eta_1\eta^*_2+\eta_2\eta^*_1 ) + \eta^*_2 (|\eta_1|^2-|\eta_2|^2)  ] + \text{c.c}  =  e^{i\gamma'} \chi' (\eta^*_1 N_2 - \eta^*_2 N_1 ) + \text{c.c} .
\end{gather}
with $\gamma'$ another arbitrary phase factor. Since the order parameters $\chi$ ($\chi'$) couple to the nematic superconductor we take the general nematic solution $(\eta_1,\eta_2) = \eta_0  ( \cos\theta, \sin\theta)$ and substitute it in $F[\chi,\eta_i] $ ($F[\chi',\eta_i] $). Taking in addition $\chi = \chi_0 e^{i \phi}$ (similarly for $\chi'$) this yields 
\begin{gather}
F[\chi_0,\eta_0] =  \chi_0  \cos (\gamma + \phi ) \eta^3_0 \cos 3\theta, \qquad F[\chi'_0,\eta_0] =  \chi'_0  \cos (\gamma' + \phi' ) \eta^3_0 \sin 3\theta . \label{eq:glmixing2}
\end{gather}
In case of $\chi$ with $A_{1u}$ symmetry, this implies that the coupling vanishes when $\theta = \pi/2 + n 2\pi/6$ ($n$ integer). This will be important below when gap structures are considered. A non-vanishing coupling means that new pairing functions are allowed to mix in by symmetry. In this particular case of $E_{2u}$ superconductors coupled to $A_{1u}$ pairing, the coupling vanishes as a result of a mirror symmetry of nematic superconductors with $\theta = \pi/2 + n 2\pi/6$.

\subsection{The case of tetragonal symmetry}

The work presented in the main text is focused on the hexagonal and trigonal crystal systems. It is worth commenting on the case of tetragonal symmetry. In case of tetragonal symmetry there is an odd-parity two-component pairing channel, given by $E_u$, but as a result of the product decomposition $E_u \times E_u = A_{1g}+A_{2g}+B_{1g}+B_{2g}$ there is no degenerate (two-component) subsidiary order. The GL theory contains an additional fourth order contribution and takes the form
\begin{gather}
F = A(T- T_c)( |\eta_1|^2 +|\eta_2|^2 )+B_{1} ( |\eta_1|^2 +|\eta_2|^2 )^2   + B_{2} | \eta_1^*\eta_2 - \eta_1\eta^*_2 |^2 + B_3(  \eta_1^*\eta_2 + \eta_1\eta^*_2 )^2 \label{eq:fetetra}.
\end{gather}
The solutions of the GL theory are given by $(1,\pm i)$ ($B_2< B_3$ with at least $B_2<0$), $(1,\pm 1)$ ($B_2> B_3$ with at least $B_3<0$), and $\{ (1, 0),(0,1) \}$ ($B_2, B_3 > 0$). As a result, whether real or imaginary (i.e., chiral) superpositions are selected depends on two GL parameters instead of one, as was the case for hexagonal ($E_{1u}$ and $E_{2u}$) and trigonal ($E_u$) symmetry. 

It is worth commenting on the two non-degenerate subsidiary order representations $B_{1g}$ and $B_{1g}$. The absence of a two-component representation describing subsidiary order quadratic in the primary fields implies that the continuous degeneracy, which is present for hexagonal and trigonal crystal systems and associated with nematic solutions, is already lifted at fourth order. In addition, we observe that within each subsidiary order channel, labeled by $B_{1g}$ and $B_{2g}$, the ground states have a twofold ($\mathbb{Z}_2$) degeneracy, whereas in case of the point groups $D_{3d}$ and $D_{6h}$ the degeneracy in the $E_g$ channel is continuous at fourth order labeled by the angle $\theta$. When crystal anisotropy is taken into account at sixth order, the degeneracy is reduced to a threefold degeneracy ($\mathbb{Z}_3$).

\subsection{Spin-rotation invariant triplet superconductors}

In the main text we compare our results for spin-orbit coupled superconductors with superconductors in which spin-orbit coupling is absent or vanishingly weak. The two-component pairing in this case was defined in Eq.~\eqref{eq:pairnosoc}. With full spin-rotation $SU(2)$ invariance the order parameters are given by $(\vec{\xi}_1,\vec{\xi}_2)$. The free energy density up to fourth order is given by
\begin{gather} 
F[\vec{\xi}_1,\vec{\xi}_2] =  A(T-T_c)(|\vec{\xi}_1|^2 + |\vec{\xi}_2|^2 ) + B_1(|\vec{\xi}_1|^2 + |\vec{\xi}_2|^2 )^2  + B_2|\vec{\xi}^*_1\times \vec{\xi}_1+\vec{\xi}^*_2\times \vec{\xi}_2|^2  + B_3 |\vec{\xi}^*_1\cdot\vec{\xi}_2 - \vec{\xi}^*_2\cdot\vec{\xi}_1 |^2 +\nonumber \\
 B_4 (\vec{\xi}^*_1\times\vec{\xi}_2 - \vec{\xi}^*_2\times\vec{\xi}_1 )^2 +  B_5(N^2_1 + N^2_2) + B_6( | \vec{N}_1|^2 + | \vec{N}_2 |^2 ) \label{eq:fenosoc} 
\end{gather}
The terms appearing in the free energy can be obtained by forming order parameter products $\xi^{a*}_i\xi^b_j$ and contracting with crystal symmetry and $SU(2)$ tensors. For instance, contracting with $(\tau_{y})_{ij}\delta^{ab}$ gives $\vec{\xi}^*_1\cdot\vec{\xi}_2 - \vec{\xi}^*_2\cdot\vec{\xi}_1$. Instead, contracting with $(\tau_{y})_{ij}\epsilon^{abc}$ gives $\vec{\xi}^*_1\times\vec{\xi}_2 - \vec{\xi}^*_2\times\vec{\xi}_1$. All of these contractions correspond to distinct subsidiary orders. 

For instance, if $B_3 < 0 $, the $\Theta$-broken chiral solution given by $(\vec{\xi}_1,\vec{\xi}_2) = \xi_0 \hat{n}(1,\pm i ) $ is favored; the direction of the $d$-vector $\hat n$ is free to rotate. The resulting superconducting state is 
the analog of two-dimensional superfluid $^3$He in the $A$ phase. The $\Theta$-invariant helical superconductor with the solution $(\vec{\xi}_1,\vec{\xi}_2) = \xi_0 (\hat{x},\hat{y}) $ is favored by $B_4< 0$. Both the chiral superconductor and the helical superconductor are rotationally invariant.

In contrast, a second class of superconducting states break rotational symmetry of the crystal and lead to a subsidiary nematic order.
Two types of anisotropic pairing can be distinguished, with scalar $N_{1,2}$ and vector $\vec{N}_{1,2}$ nematic order respectively. In terms of the contractions the scalar nematic orders are given by $N_1 = \xi^{a*}_{i}(\tau_z)_{ij}\xi^{a}_{j} = \xi^{a*}_{i}(\tau_z)_{ij}\delta^{ab}\xi^{b}_{j}$ and $N_2 = \xi^{a*}_{i}(\tau_x)_{ij}\xi^{a}_{j}= \xi^{a*}_{i}(\tau_x)_{ij}\delta^{ab}\xi^{b}_{j}$. Explicitly they take the form
\begin{gather*} 
( N_1,N_2 ) = (|\vec{\xi}_1|^2 - |\vec{\xi}_2|^2 , \vec{\xi}^*_1\cdot\vec{\xi}_2 + \vec{\xi}^*_2\cdot\vec{\xi}_1). \label{eq:nemscal}
\end{gather*}
Instead, the vector nematic orders are given by $N^c_1 =\epsilon^{cab} \xi^{a*}_{i}(\tau_z)_{ij}\xi^{b}_{j}$ and $N^c_2 = \epsilon^{cab}\xi^{a*}_{i}(\tau_x)_{ij}\xi^{b}_{j}$, which is explicitly expressed as
\begin{gather*} 
( \vec{N}_1 , \vec{N}_2 ) =  ( \vec{\xi}^*_1\times \vec{\xi}_1- \vec{\xi}^*_2\times \vec{\xi}_2 , \vec{\xi}^*_1\times\vec{\xi}_2 + \vec{\xi}^*_2\times\vec{\xi}_1).  \label{eq:nemvec}
\end{gather*} 
Under time-reversal symmetry $\Theta$ one has $\xi^{a}_{i} \rightarrow \xi^{a*}_{i}$ and therefore scalar nematic order is time-reversal even, whereas vector nematic order is time-reversal odd. 

In the main text we have argued, and below we will show, that the chiral and helical superconductors, expressed as $(\vec{\xi}_1,\vec{\xi}_2) = \xi_0 \hat{n}(1,\pm i ) $ and $(\vec{\xi}_1,\vec{\xi}_2) = \xi_0 (\hat{x},\hat{y}) $ are always favored over other solutions, and at the level of our analysis remain degenerate solutions. It is important to stress that this conclusion does not depend on details such as Fermi surface shape. This conclusion only depends on and directly follows from symmetry: for any representation corresponding to two-component pairing, described by orbital form factors $g_1(\vec{k})$ and $g_2(\vec{k})$ transforming as the twofold representation, the chiral and helical superconductors can be shown to have lower energy within weak-coupling BCS (see below). We note that this is also true in two dimensions and for symmetry groups $D_6$ and $D_3$ in case of pairing representations $E_{1}$ and $E$.

\section{Weak-coupling calculation of coefficients\label{sec:bcs}}

Within weak-coupling BCS theory the GL coefficients can be microscopically calculated by evaluating Feynman diagrams. Here we describe in more detail how the expression in the main text were obtained. 

The quasiparticle part of the mean-field free energy can be expressed as 
\begin{gather}
-\frac{1}{\beta} \text{Tr}\, \ln \mathcal{G}^{-1} = F_0
-\frac{1}{\beta} \text{Tr}\, \ln ( 1- \mathcal{G}_0\Sigma), \label{eq:trace}
\end{gather}
where $\beta$ is inverse temperature, $\mathcal{G}$ is the mean-field Gorkov Green's function and $\mathcal{G}_0$ is the normal state Green's function given by
\begin{gather}
\mathcal{G}_0 = \begin{pmatrix}G_+ &  \\
   & G_- \end{pmatrix}= \begin{pmatrix} G_+(\vec{k},i\omega)  &
   \\    & - G_+(\vec{k},-i\omega)
 \end{pmatrix}.
\end{gather}
We have used the notation $G_\pm$ to denote the electron ($+$) and hole ($-$) Green's functions. These are expressed in the basis defined by Eq.~\eqref{eq:nambu}. We have used the notation $G_\pm$ to denote the electron ($+$) and hole ($-$) Green's functions. The mean-field self-energy $\Sigma$ contains the superconducting order parameters and depends on the pairing channel. In case of spin-orbit coupling it is given by
\begin{gather}
\Sigma = \begin{pmatrix}  & \Delta \\
  \Delta^\dagger &  \end{pmatrix}= \sum_m \begin{pmatrix}  & \eta_m\Delta_m(\vec{k}) \\
  \eta^*_m\Delta^\dagger_m(\vec{k})  &  \end{pmatrix} . \label{eq:selfen}
\end{gather}
The mean-field self energy in the absence of spin-orbit coupling (i.e., with spin-rotation invariance) is given by
\begin{gather*}
\Sigma = \begin{pmatrix}  & \Delta \\
  \Delta^\dagger &  \end{pmatrix}=  \begin{pmatrix}  & \xi^i_mg_m  \sigma^i \\
 (\xi^{i}_mg_m )^* \sigma^i  &  \end{pmatrix} ,
\end{gather*}
where the sum over repeated indices is implied.

The $n>2 $ terms of the GL free energy are obtained by expanding~\eqref{eq:trace} to a given order in the order parameter fields. This is a convenient way of generating the Feynman diagrams needed to calculate the expansion coefficients. The fourth and sixth order contributions are given by
\begin{gather} 
F^{(4)} + F^{(6)} = \frac{1}{4\beta} \trace{(\mathcal{G}_0\Sigma )^4} + \frac{1}{6\beta} \trace{(\mathcal{G}_0\Sigma )^6}. \label{eq:4thorder}
\end{gather}
The trace should be read as an integral over momenta and sum over frequencies, in addition to a trace over particle-hole and internal spin space: $\text{Tr} \equiv \sum_\omega\int_{\vec{k}} \text{tr} $. 

Next, we define the momentum dependent form factor functions $g_i(\vec{k})$ ($i=1,2,3$) to transform according to representations of the crystal symmetry group. Specifically, $(g_1,g_2)$ transform according to $E_u$ and $E_{1u}$, and $g_3$ transforms as $A_{2u}$. To leading order they are given by the $L=1$ ($p$-wave) orbital form factors as 
\begin{gather}  
(g_1, g_2, g_3) = (k_x, k_y, k_z), \quad g_\pm = k_x \pm i k_y
\end{gather}
The Feynman diagram integrals will contain products of these orbital form factor functions. Using the expression for the mean-field self energy $\Sigma$ the fourth order contribution to the free energy takes the form
\begin{gather}  
F^{(4)}  = \frac{1}{2} T \sum_\omega \int_{\vec{k}} G^2_+ G^2_-  \text{tr}[ \Delta\Delta^\dagger \Delta\Delta^\dagger] \label{eq:integral}
\end{gather}
To proceed we have to choose a pairing channel and specify the two-component pairing functions $\Delta_m(\vec{k})$ corresponding to that channel. Before we present the results for the cases considered in the main text, we adopt the convention of the main text and abbreviate the integral over Green's functions and form factors as $\order{\ldots}$. As an example, $\order{g^2_1 g^2_2}$ is a short hand notation for
\begin{gather}  
\order{g^2_1 g^2_2} \equiv T \sum_{\omega} \int_{\vec{k}} \; G^2_+ G^2_-  g^2_1(\vec{k}) g^2_2(\vec{k}).
\end{gather}

\subsection{Spin-orbit coupled superconductors}

We first consider the pairing functions listed in Table I of the main text, which are leading order crystal harmonic functions, and then consider the general cases. We will start with the hexagonal group $D_{6h}$, which has two distinct two-component odd-parity channels, and then turn to trigonal symmetry ($D_{3d}$). 

Lowest order pairing functions in the $E_{2u}$ channel are given by
\begin{gather}  
\Delta_1(\vec{k}) = g_1\sigma_x - g_2\sigma_y , \quad  \Delta_2(\vec{k}) =g_1\sigma_y + g_2\sigma_x \label{eq:pair1}
\end{gather}
($F^c_{1,2}(\vec{k})$ of the main text). Substituting this into Eq.~\eqref{eq:integral} we find that the GL coefficients $B_{1,2}$ of Eq.~\eqref{eq:fehexatrig} are given by a single integral 
\begin{gather}  
B_1 = B_2 = \order{g^2_+g^2_- },
\end{gather}
in terms of the form factors $g_{1,2}$, which to leading order are $k_{x,y}$. For a stable free energy $B_1>0$, and therefore we conclude that the nematic superconductor is favored in weak-coupling. 

Next, we consider the pairing functions in the $E_{1u}$ channel
\begin{gather}  
\Delta_1(\vec{k}) = \lambda_a g_1\sigma_z + \lambda_bg_3\sigma_x , \quad  \Delta_2(\vec{k}) =\lambda_a g_2\sigma_z + \lambda_bg_3 \sigma_y,
\end{gather}
which are linear combinations of the $F^{a,b}_{1,2}(\vec{k})$ pairings of the main text with coefficients $\lambda_a,\lambda_b$. We find that the coefficients $B_i$ can be expressed in terms of two integrals $I_i$ given by
\begin{gather}  
I_1=  \lambda_a^4\order{g^2_+g^2_- }/8, \quad I_2 =  \lambda_b^2\order{g^2_3 (\lambda_a^2g_+ g_-+\lambda_b^2g^2_3) }. \label{eq:e1u}
\end{gather}
Explicitly, the coefficients $B_i$ are expressed as $B_1 = 3I_1 + I_2$ and $B_2 = -I_1 + I_2$. In the limit $\lambda_a=0$ we have that $I_1=0$ and $\lambda_b=0$ corresponds to $I_2=0$. 

In case of the symmetry group $D_{3d}$ there is only a single two-component odd-parity channel: $E_u$. We therefore write the pairing components as arbitrary linear combinations of the basis functions $F^{a,b,c}_{1,2}$ of the main text as
\begin{gather}  
\Delta_1(\vec{k}) = \lambda_a g_1\sigma_z + (\lambda_bg_3 + \lambda_cg_2)\sigma_x + \lambda_cg_1\sigma_y, \quad
 \Delta_2(\vec{k}) =\lambda_ag_2\sigma_z + (\lambda_bg_3 - \lambda_cg_2)\sigma_y + \lambda_cg_1 \sigma_x, \label{eq:pairingeu}
\end{gather}
with coefficients $\lambda_a,\lambda_b,\lambda_c$. We find that the coefficients $B_i$ are expressed in terms of two integrals $I_i$ as
\begin{gather}  
I_1=  \lambda_a^4\order{g^2_+g^2_- }/8+ 2\lambda_b^2\lambda_c^2\order{g^2_3g_+g_- }, \nonumber \\ 
I_2 = \lambda_c^2(\lambda_a^2 +\lambda_c^2)\order{g^2_+ g^2_-} +\lambda_b^2(\lambda_a^2-2\lambda_c^2)\order{g_+ g_-g^2_3} + \lambda_b^4\order{g^4_3} +i\frac{\lambda_a^2\lambda_b\lambda_c}{2}\order{g_3(g^3_+ - g^3_-)} . \label{eq:eupair}
\end{gather}
Explicitly, the coefficients $B_i$ are expressed as $B_1 = 3I_1 + I_2$ and $B_2 = -I_1 + I_2$. In the limit of $c=0$ we recover the result of hexagonal $E_{1u}$ pairing channel.  

As a next step, we consider an example of pairing functions which are superpositions of two degenerate basis functions beyond lowest order in crystal harmonics. We take the example of a $3D$ hexagonal crystal, where the pairing functions of Eq.~\eqref{eq:pair1} are degenerate with $(k_x^2-k_y^2,2k_xk_y)k_z\sigma_z$ in the $E_{2u}$ channel. The pairing is then a linear combination of both basis functions, given by 
\begin{gather}  
\Delta_1(\vec{k}) = \lambda_a(g_1\sigma_x - g_2\sigma_y)+\lambda_b k_z(k_x^2-k_y^2)\sigma_z , \quad
 \Delta_2(\vec{k}) =\lambda_a( g_1\sigma_y + g_2\sigma_x)+\lambda_b 2 k_zk_xk_y\sigma_z 
\end{gather}
Again substituting this into Eq.~\eqref{eq:integral} we find the two integrals $I_i$ to be given by
\begin{gather}  
I_1= \lambda_b^4 \order{h^2_+ h^2_-}/8, \quad I_2 =   \lambda_a^2\order{ g_+g_-(\lambda_a^2g_+g_-+\lambda_b^2h_+ h_-)}
\end{gather}
where $(h_1,h_2) = ( k_z\{k_x^2-k_y^2\}, 2k_zk_xk_y)$ and $g_\pm = g_1 \pm i g_2$. The coefficients $B_{i}$ are expressed in the $I_i$ integrals as before. We therefore find that admixture of higher crystal harmonics, in this case $(h_1,h_2)$, leads to a finite $I_1$, i.e., the integral which signals the tendency towards  chiral superconductivity.

\subsection{The general case}

We now consider the general case in which no specific assumptions regarding the form of the pairing functions are made. The pairing functions will be general linear combinations of degenerate crystal harmonics, expressed in Eq.~\eqref{eq:expand}, and we write the two components as
\begin{gather}  
\Delta_1(\vec{k}) = \vec{d}_1 \cdot \vec{\sigma} , \quad  \Delta_2(\vec{k}) =\vec{d}_2 \cdot \vec{\sigma}  ,
\end{gather}
bearing in mind that the $\vec{d}$-vectors depend on momentum. The pairing takes the form $\Delta = \eta_1 \Delta_1(\vec{k}) + \eta_2 \Delta_2(\vec{k}) $, where $\eta_{1,2}$ are the complex order parameters. In order to evaluate Eq.~\eqref{eq:integral} we first calculate $\Delta\Delta^\dagger $ and find (suppressing momentum dependence)
\begin{gather}  
\Delta\Delta^\dagger  = d^2_1|\eta_1|^2 + d^2_2|\eta_2|^2 + \eta_1\eta^*_2(\vec{d}_1 \cdot \vec{d}_2  +i \vec{d}_1 \times\vec{d}_2  \cdot \vec{\sigma} )  + \eta_2\eta^*_1(\vec{d}_1 \cdot \vec{d}_2  +i \vec{d}_2 \times\vec{d}_1  \cdot \vec{\sigma} ) 
\end{gather}
Then, calculating $\Delta\Delta^\dagger \Delta\Delta^\dagger$ and taking the trace we find after some straightforward rewriting
\begin{gather}  
\frac{1}{2} \text{tr}[ \Delta\Delta^\dagger \Delta\Delta^\dagger]  = d^4_1|\eta_1|^4 + d^4_2|\eta_2|^4 + 2 |\eta_1|^2|\eta_2|^2[(\vec{d}_1 \times\vec{d}_2)^2 + 3(\vec{d}_1 \cdot \vec{d}_2)^2 ] + 
[ (\vec{d}_1 \times\vec{d}_2)^2  -  (\vec{d}_1 \cdot \vec{d}_2)^2]  |\eta_1\eta^*_2 - \eta_2\eta^*_1|^2  \label{eq:4trace}
\end{gather}
Note that on the right hand side we have neglected all terms which cannot appear in the free energy (e.g., terms such as $\eta_1\eta^*_2|\eta_1|^2$, which must vanish when performing the momentum integration). Now, combining this with Eq.~\eqref{eq:integral} we obtain the expressions discussed in the main text. Specifically, we have $B_2 = \order{ (\vec{d}_1 \times\vec{d}_2)^2} - \order{(\vec{d}_1 \cdot \vec{d}_2)^2}$ with a positive contribution coming from $\vec{d}_1(\vec{k}) \times\vec{d}_2(\vec{k})$ (reinstating momentum dependence) and a negative contribution from $\vec{d}_1(\vec{k}) \cdot \vec{d}_2(\vec{k})$.  

The next step is to explicitly use symmetry to combine the remaining terms into $B_1 (|\eta_1|^2+|\eta_2|^2 )^2$. We consider the following object 
\begin{gather}  
 (\vec{d}_1+i\vec{d}_2)^4 + (\vec{d}_1-i\vec{d}_2)^4 = 2(d^4_1+d^4_2 ) - 4d^2_1d^2_2 -8 (\vec{d}_1 \cdot \vec{d}_2)^2,
\end{gather}
which allows us to write
\begin{gather}  
 \frac{1}{2}(d^4_1+d^4_2 )  = d^2_1d^2_2 +2 (\vec{d}_1 \cdot \vec{d}_2)^2 +\frac{1}{4} (\vec{d}_1+i\vec{d}_2)^4 + \frac{1}{4}(\vec{d}_1-i\vec{d}_2)^4 = (\vec{d}_1 \times\vec{d}_2)^2 + 3(\vec{d}_1 \cdot \vec{d}_2)^2 +\frac{1}{4} (\vec{d}_1+i\vec{d}_2)^4 + \frac{1}{4}(\vec{d}_1-i\vec{d}_2)^4. 
\end{gather}
Because of symmetry we can substitute the coefficients of $|\eta_{1,2}|^4$ in Eq.~\eqref{eq:4trace}, given by $d^4_{1,2}$, both by $(d^4_{1}+d^4_{2})/2$. Then, this implies for the first three terms of the trace of Eq.~\eqref{eq:4trace}
\begin{gather}  
 d^4_1|\eta_1|^4 + d^4_2|\eta_2|^4 + 2 |\eta_1|^2|\eta_2|^2[(\vec{d}_1 \times\vec{d}_2)^2 + 3(\vec{d}_1 \cdot \vec{d}_2)^2 ]  \simeq ((\vec{d}_1 \times\vec{d}_2)^2 + 3(\vec{d}_1 \cdot \vec{d}_2)^2 )(|\eta_1|^2 + |\eta_2|^2)^2  \nonumber \\
 + \frac{1}{4}[ (\vec{d}_1+i\vec{d}_2)^4+(\vec{d}_1-i\vec{d}_2)^4]( |\eta_1|^4+|\eta_2|^4) \label{eq:terms}
\end{gather}
In case of hexagonal and trigonal systems the second term of~\eqref{eq:terms} must vanish for reasons of symmetry, and we are left with the first term. This shows that in the general case we have $B_1 = 3I_1 + I_2$ and $B_2 = -I_1 + I_2$ with
\begin{gather}  
I_1 = \order{(\vec{d}_1 \cdot \vec{d}_2)^2}, \qquad I_2 = \order{(\vec{d}_1 \times \vec{d}_2)^2}.
\end{gather}
As a result of the relation $B_1 = 3I_1 + I_2$ and the underlying assumption $B_1>0$ (requirement of a stable free energy) our conclusion that $I_1$ favors chiral superconductivity and $I_2$ favors nematic superconductivity, which we presented in the main text, follows.

We can now make the difference with tetragonal symmetry (see also Eq.~\eqref{eq:fetetra}) more precise. In case of tetragonal the integral
\begin{gather}  
\order{ (\vec{d}_1+i\vec{d}_2)^4+(\vec{d}_1-i\vec{d}_2)^4}
\end{gather}
generally does not vanish, leading to an additional term in the free energy with coefficient $B_3$ in Eq.~\eqref{eq:fetetra}. 

\subsection{Spin-rotation invariant superconductors}

In the same way as the spin-orbit coupled cases considered so far, we can derive the weak-coupling GL coefficients for the case when spin-orbit coupling is absent. Using Eq.~\eqref{eq:integral} and working out the traces we find the free energy given in the main text with the coefficients 
\begin{gather}  
\frac{1}{4} \order{g^2_+g^2_- } = B_1= B_2= 2B_5= 2B_6.
\end{gather}
Other terms are absent, implying that $B_3=B_4=0$. Except for the latter, all coefficients are positive and we thus conclude that the chiral and helical superconductor are favored. The two remain degenerate, however. (Note that this result only holds for hexagonal and trigonal symmetry, which we mainly focus on.)


%
%
%

\subsection{Calculation of integrals}

We now make our analysis more quantitative and calculate the integrals needed to evaluate the GL coefficients. Recall that the particle and hole Green's functions are defined as  $G_\pm = (i\omega \mp E(\vec{k}))^{-1} $ where $E(\vec{k})  = \epsilon(\vec{k}) - \mu$. The fourth order GL terms are all expressed as a single integral type of the general form $\order{A(\hat{k})}$, where $A(\hat{k})$ is some product of momentum form factor functions. It is given by Eq.~\eqref{eq:integral} and reads
\begin{gather}  
\order{A(\hat{k})} \equiv T \sum_{\omega} \int_{\vec{k}} \; G^2_+ G^2_-  A(\hat{k}) = T \sum_n \int \frac{d^3k}{(2\pi)^3} \frac{A(\hat{k})}{(\omega_n^2+ E^2)^2} = T \sum_n \int \frac{dk k^2}{(2\pi)^3} \frac{1}{(\omega_n^2+ E^2)^2} \int d\Omega_{\hat{k}} A(\hat{k})
\end{gather}
Here we have made the assumption that $\epsilon(\vec{k})$ corresponds to a spherically symmetric dispersion. In particular, the Fermi surface is assumed to be a perfect sphere. We find that the integral evaluates to 
\begin{gather}  
\order{A(\hat{k})} = T \sum_n \int \frac{dk k^2}{(2\pi)^3} \frac{1}{(\omega_n^2+ E^2)^2} \int d\Omega_{\hat{k}} A(\hat{k}) = \frac{N(\epsilon_F)}{2(\pi T_c)^2}  \frac{7\zeta(3)}{4} \int \frac{d\Omega_{\hat{k}}}{4\pi} A(\hat{k}),
\end{gather}
where $N(\epsilon_F)$ is the density of states per spin projection at the Fermi level. What remains is the integral of the form factors $A(\hat{k})$ over the Fermi surface ($\Omega_{\hat{k}}$ is the solid angle). Since we are assuming a perfectly spherical Fermi surface, we can write $\vec{k} = k_F(\cos\varphi\sin\theta,\sin\varphi\sin\theta,\cos\theta)$ or $\hat{k} = (\cos\varphi\sin\theta,\sin\varphi\sin\theta,\cos\theta)$. The latter is then substituted for $A(\hat{k})$, which allows to evaluate the Fermi surface integrals. 

We now consider the specific example of $E_u$ pairing in a trigonal crystal. The integrals are given in Eq.~\eqref{eq:eupair}, and we therefore need $\order{g^2_+g^2_- }$, $\order{g^2_3g_+ g_- }$ and $\order{g^4_3 }$, with $g_\pm = k_x \pm ik_y$ and $g_3 = k_z$:
\begin{gather}  
 \int \frac{d\Omega_{\hat{k}}}{4\pi} \hat{k}^4_z  = \frac{1}{2} \int_0^\pi d\theta \sin\theta \cos^4\theta = \frac{1}{5} \nonumber \\
  \int \frac{d\Omega_{\hat{k}}}{4\pi}  (\hat{k}^2_x +\hat{k}^2_y)^2    =  \frac{1}{2} \int_0^\pi d\theta \sin^5\theta =  \frac{8}{15} \nonumber \\
  \int \frac{d\Omega_{\hat{k}}}{4\pi}  (\hat{k}^2_x +\hat{k}^2_y)\hat{k}^2_z  =  \frac{1}{2} \int_0^\pi d\theta \sin^3\theta \cos^2\theta =  \frac{2}{15}
\end{gather}
We define $x = \lambda_a/\lambda_c$ and $y = \lambda_b/\lambda_c$. We can then express the GL coefficient $B_2$, given by $B_2 = -I_1+I_2$ in terms of $x,y$ and find $B_2 \propto -(x^4 + 4y^2) + 8(x^2+1) + 2y^2(x^2-2)+3y^4$. The resulting phase diagram as function of $x,y$ is shown in Fig. 1 of the main text.
 
\section{Quasiparticle gap structures of odd-parity superconductors\label{sec:gaps}}

The quasiparticle energies in the presence of pairing potential $\Delta(\vec{k}) $ depend on the structure of $\Delta^\dagger\Delta$. Writing the pairing in terms of a $\vec{d}$-vector as $\Delta(\vec{k}) = \vec{d}(\vec{k})\cdot \vec{\sigma}$, where $\vec{d}(\vec{k})$ collects all momentum dependent components, one finds $\Delta^\dagger\Delta = |\vec{d}(\vec{k})|^2 \sigma_0 +i \vec{d}^*\times \vec{d}\cdot \vec{\sigma}$. If $\vec{d}^*\times \vec{d} = 0$ the pairing is unitary, otherwise the pairing is non-unitary. The quasiparticle energies $\mathcal{E}_\pm(\vec{k}) $ are given by
\begin{gather} 
\mathcal{E}_\pm(\vec{k}) = \sqrt{ (\varepsilon - \mu)^2 + |\vec{d}(\vec{k})|^2  \pm | \vec{d}^*(\vec{k})\times \vec{d}(\vec{k})|}.
\end{gather} 
This expression applies to both unitary and non-unitary pairing and in case of the former it clearly reduces to $\mathcal{E}_\pm(\vec{k}) = \sqrt{ (\varepsilon - \mu)^2 + |\vec{d}(\vec{k})|^2}$. We now present the gap structures for the superconducting states considered in this work, starting with the case of no spin-orbit coupling. 

\begin{figure}
\includegraphics[width=0.25\textwidth]{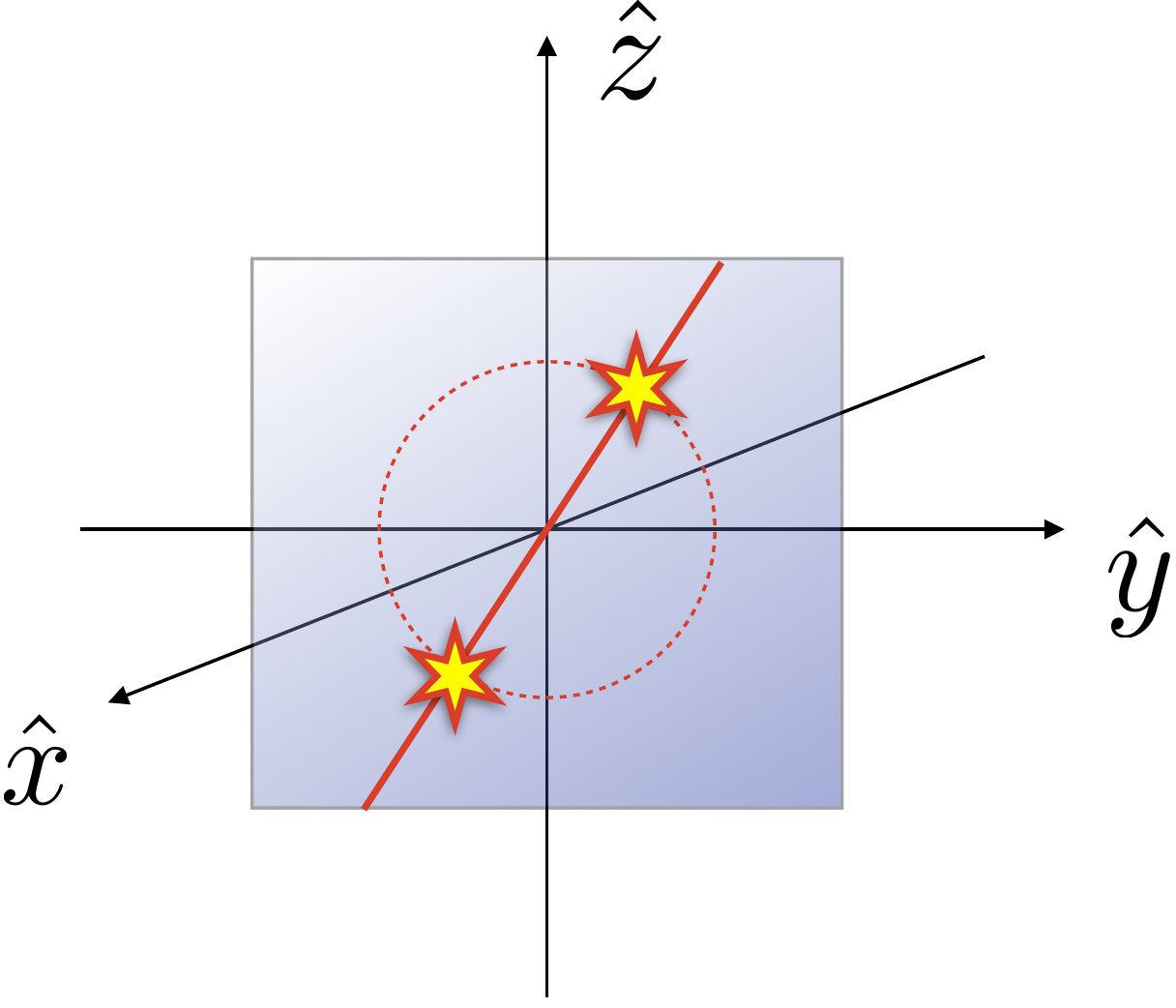}
\caption{\label{fig:mirror} For a mirror symmetry $M_{yz}$ in the $y-z$ plane the $\vec d$-vector $\vec{d}(k_x=0,k_y,k_z)$ must be parallel to the mirror plane normal, i.e., $d^{y,z}(k_x=0,k_y,k_z) = 0$. The remaining function $d^{x}(k_x=0,k_y,k_z)$ is a function of two variables and in general has a line (or dimension 1 manifold) of zeros in the $y-z$ plane (solid red line). Where the line intersects the Fermi surface (dashed red circle) the quasiparticle spectrum will be gapless, exhibiting spin-degenerate point nodes. Note that this argument assumes $\vec d $ is real ($\Theta$-invariant pairing). }
\end{figure}

Before we present the gap structures we briefly review the protection of point node degeneracies of spin-orbit coupled $\Theta$-invariant triplet superconductors with real $\vec d$-vector by a mirror reflection symmetry. In the presence of spin-orbit coupling the $\vec d$-vector is generally a three component function of momentum, i.e., $\vec{d}(\vec{k})$, since crystalline symmetry is insufficient to make a component of the $\vec d$-vector vanish everywhere in $k$-space. Satisfying $\vec{d}(\vec{k})=0$ on the Fermi surface, which amounts to three equations as function of \emph{two} variables (parametrizing the Fermi surface), is vanishingly improbable. This is different in the presence of a mirror reflection symmetry. Mirror symmetry forces the $\vec{d}(\vec{k})$ vector to be normal to the mirror plane \emph{on the mirror plane}. Vanishing of the $\vec d$-vector on the mirror plane than requires satisfying a single equation of two variables, which is generally possible on a dimension-1 manifold (e.g., a line). The intersection of the $1D$ manifold with the Fermi surface results in point nodes (see also Fig.~\ref{fig:mirror}).

\subsection{Without spin-orbit coupling}

The pairing matrices of the chiral and helical superconductors are given by
\begin{gather}
\text{chiral :}\quad \Delta(\vec{k}) = \xi_0 \sigma_z(k_x \pm ik_y) \\
\text{helical :}\quad \Delta(\vec{k}) =  \xi_0(k_x  \sigma_x + k_y  \sigma_y) 
\end{gather}
Note that we have chosen a particular $\vec d$-vector, but all $\vec d$-vectors obtained by rotations in spin space are equivalent. Both the chiral and helical superconductors are examples of unitary superconducting states, and one simply finds $\frac{1}{2}\text{Tr} \Delta^\dagger\Delta= \xi^2_0(k^2_x + k^2_y) $ in both cases. In $2D$ the chiral and helical superconductor are fully gapped. In $3D$ the quasiparticle energy spectrum corresponding to $\frac{1}{2}\text{Tr}  \Delta^\dagger\Delta= \xi^2_0(k^2_x + k^2_y) $ shows spin-degenerate nodes along the $z$ direction. In $3D$, however, other pairings with the same symmetry can mix in, which in case of the helical superconductor is given by $k_z \hat{z}$. Such admixture leads to a fully gapped helical superconductor, whereas the chiral superconductor remains nodal.

The scalar nematic superconductor has unitary pairing matrix $\Delta(\vec{k}) = \xi_0 \sigma_z(\cos\theta k_x + \sin\theta k_y)$ and gives rise to the gap structure 
\begin{gather*}
 \frac{1}{2}\text{Tr} \Delta^\dagger\Delta = \xi^2_0[k^2_x +k_y^2 + \cos 2\theta(k^2_x - k^2_y) + 2\sin 2\theta \,k_xk_y] /2,
\end{gather*}
which has line nodes, the location of which depends on $\theta$. For instance, $\theta =0 $ leads to a line node in the $y-z$ plane. Finally, the vector nematic state, an example of which was given in the main text, is a non-unitary state and the pairing matrix is given by $\Delta(\vec{k}) = \eta_0( k_x + k_y) \sigma_\pm $. In this case only one (pseudo)spin species develops an (anisotropic) nodal pairing gap.

\subsection{Spin-orbit coupled pairing}

We now discuss the gap structures of spin-orbit coupled superconductors. First, we present the quasiparticle gap structures of superconductors with symmetry group $D_{6h}$. In the hexagonal crystal system two two-component representations are distinguished: $E_{2u}$ and $E_{1u}$. We then proceed to $E_u$ pairing in trigonal crystals. 

\subsubsection{Hexagonal crystal symmetry: $E_{2u}$ pairing}

Consider first $E_{2u}$ pairing symmetry. In the leading order $p$-wave harmonic expansion the two pairing partners are given by $(k_x \sigma_x - k_y \sigma_y,k_x \sigma_y + k_y \sigma_x) $. The nematic superconductor $(\eta_1,\eta_2) =  \eta_0(1,0)$ has pairing matrix
\begin{gather}
 \Delta_1(\vec{k}) = \eta_0 (k_x \sigma_x - k_y \sigma_y),
\end{gather}
and a gap structure is given by $|\vec{d}|^2 = \eta^2_0 (k_x^2+k_y^2)$, which implies spin-degenerate point nodes along the $z$ axis. It has a full pairing gap in the $x-y$ plane. The same holds for the $(\eta_1,\eta_2) =  \eta_0(0,1)$ superconductor. The chiral superconductor, defined by the solution $(\eta_1,\eta_2) =  \eta_0(1,i)$, has pairing matrix
\begin{gather}
 \Delta_+(\vec{k}) = \eta_0 k_+ \sigma_+. 
\end{gather}
The chiral superconductor is a non-unitary pairing state which has a pairing gap for only one of the pseudospin species, while the other does not develop pairing. The gap of the paired species has nodes along the $z$ direction, resulting in a spin-degenerate point node along $z$ for this pseudospin species. 

The accidental rotational symmetry of the quasiparticle spectrum of the nematic superconductor is an indication that this is not the most general pairing state of a $(\eta_1,\eta_2) =  \eta_0(1,0)$ nematic superconductor in the $E_{2u}$ channel. In Sec.~\ref{sec:oddparity} we have discussed how general pairing functions are linear combinations of an infinite set of degenerate basis functions. Specifically, based on Eq.~\eqref{eq:expand} one can choose a more general $\eta_2=0$ nematic state composed of $E_{2u}$ functions. Very specific choices, such as the leading order crystal harmonic, may have properties which are not generic for the $\eta_2=0$ nematic superconductor. In particular, including higher harmonics leads to a dependence of the gap on the angle $\theta_k$ (where $k_\pm = k e^{\pm i\theta_k}$). 

Similarly, a more general chiral superconducting state $(\eta_1,\eta_2) =  \eta_0(1,i)$ in the $E_{2u}$ channel can be written as an expansion in higher order crystal harmonics. Using the harmonics listed in Sec.~\ref{sec:oddparity} it is simple to show that a general chiral superconductor is a non-unitary pairing state. 

To rigorously establish which properties of the gap structures are generic and manifest we resort to symmetry arguments. We write a general pairing in the $E_{2u}$ channel as $\Delta_{1,2}(\vec{k}) = \vec{d}_{1,2}(\vec{k})\cdot \vec{\sigma}$ and first focus on the $z$ axis, i.e., $\vec{k} = (0,0,k_z)$. As a result of odd-parity we must have $ \vec{d}_{1,2}(0,0,-k_z) = - \vec{d}_{1,2}(0,0,k_z)$. In addition, since $E_{2u}$ is even under a two-fold rotation about the $z$-axis, we should have $d^{x,y}_{1,2}(0,0,k_z) = - d^{x,y}_{1,2}(0,0,k_z)$, which requires these components to be zero. Furthermore, $d^{z}_{1,2}(0,0,k_z)$ are forced to be zero by the requirement that they describe a two-component pairing representation. These arguments show that the most general linear combination of $E_{2u}$ pairing functions will give rise to spin-degenerate nodes along the $z$ direction, both for the nematic superconductors and the chiral superconductors. 

This, however, does not prove that the nodes are symmetry protected. Since both the nematic and chiral superconductors break symmetries, pairing terms originating from symmetry-distinct representations can mix as a result of the lower symmetry. Let us consider the two nematic superconductors $(\eta_1,\eta_2) = \eta_0(1,0)$ and $(\eta_1,\eta_2) = \eta_0(0,1)$. The former is even under the two-fold rotation about the $x$ axis $C'_{2x}$ and odd under the mirror reflection $x \rightarrow -x$. This implies that the pairing $k_z \hat{z}$, which transforms the same way under these symmetries, can mix in and fully gap out the spin-degenerate nodes. In contrast, the nematic superconductor $(0,1)$ transforms in the opposite way, preventing a mixing in of $k_z \hat{z}$. The mixing in of pairings with distinct symmetry can be understood from the GL theory discussed in Sec.~\ref{sec:landau}. In particular, the coupling of different representations expressed in Eqs~\eqref{eq:glmixing}  and~\eqref{eq:glmixing2} directly applies to the present case. The pairing $k_z \hat{z}$ has $A_{1u}$ symmetry and according to Eq.~\eqref{eq:glmixing2} will be maximally induced when the angle characterizing the nematic superconductor equals $\cos 3\theta=\pm 1$. When $\cos 3\theta= 0 $ no coupling occurs. 

A similar symmetry-based argument holds for the chiral superconductor. A general chiral superconductor has spin-degenerate point nodes along the $z$ axis since any pairing function in the $E_{2u}$ channel vanishes along $z$. A general chiral superconductor, however, will develop magnetic moment due to its non-unitarity, and therefore allows for a Zeeman field to couple to the spin. As a result, spin degeneracy is lifted and the degenerate point nodes are split. 

\subsubsection{Hexagonal crystal symmerty: $E_{1u}$ pairing}

Let us now proceed to the $E_{1u}$ channel. An example of a $(\eta_1,\eta_2) = \eta_0(1,0)$ nematic superconductor is given by
\begin{gather}
 \Delta_1(\vec{k}) = \eta_0 \vec d_1 \cdot \vec{\sigma}, \quad \text{with} \quad \vec d_1 =  \text{Re}\, \left[ \lambda_a k_+ \hat{z} + \lambda_bk_z \hat{r}_+   \right],
\end{gather}
in terms of three degenerate pairing functions and expansion coefficients $\lambda_{a,b}$. The gap structure takes the form
\begin{gather}
|\vec{d}|^2 = \eta^2_0 [ \lambda_a^2k^2\cos^2\theta_k +  \lambda^2_b k_z^2],
\end{gather}
where we use the same notation as before ($k^2 = k_x^2+k_y^2$). In general, the $(1,0)$ nematic superconductor will have a full pairing gap along the $z$ axis. In the $x-y$ plane spin-degenerate point nodes are present when $\cos \theta_k = 0$, i.e., along the $y$ axis. A nematic superconductor $(\eta_1,\eta_2) = \eta_0(0,1)$ with expansion coefficients $\lambda_{a,b}$ is given by
\begin{gather}
 \Delta_2(\vec{k}) = \eta_0 \vec d_2 \cdot \vec{\sigma}, \quad \text{with} \quad \vec d_2 =  \text{Im}\, \left[ \lambda_a k_+ \hat{z} + \lambda_bk_z \hat{r}_+  \right].
\end{gather}
Its gap structure is very similar to that of its partner and given by
\begin{gather}
|\vec{d}|^2 = \eta^2_0 [ \lambda_a^2k^2\sin^2\theta_k  +  \lambda_b^2 k_z^2 ]
\end{gather}
The $(0,1)$ superconductor has a full pairing gap along the $z$ axis and spin-degenerate point nodes in the $x-y$ plane given by the condition $\cos \theta_k = 0$. 

In the same way as for $E_{2u}$ pairing, the most general nematic superconductors in the $E_{1u}$ channel is obtained by taking a general expansion in crystal harmonics, following Sec.~\ref{sec:oddparity}. The generic features of the gap structures are more conveniently addressed by symmetry arguments. 
The gap structure of the nematic superconductors, in particular the spin-degenerate point nodes, can be understood from the mirror symmetries. The superconductor $(1,0)$ has a mirror symmetry $ x \rightarrow -x$, whereas the $(0,1)$ superconductor is symmetric under the mirror reflection $ y \rightarrow -y$. This implies symmetry protected Dirac nodes in the $z-y$ plane and $z-x$ plane, respectively. In addition, since the nematic superconductors in the $E_{1u}$ channel have a mirror symmetry $z \rightarrow -z$ (parity combined with twofold rotation $C_{2z}$) the Dirac nodes are in the $x-y$ plane. As a result, the superconductor $(1,0)$ has Dirac nodes along $y$ and the superconductor $(0,1)$ has Dirac nodes along $x$. For a generic nematic superconductor $(\eta_1,\eta_2)$ the only mirror symmetry is $z \rightarrow -z$, which protects point nodes in the $x-y$ plane which are not pinned to an axis. 

The chiral superconductor $(\eta_1,\eta_2) = \eta_0(1,i)$ in the $E_{1u}$ channel, with the same expansion coefficients as the nematic superconductors, has the gap structure
\begin{gather}
|\vec{d}|^2 \pm |\vec{d}^*\times \vec{d} |= \eta^2_0 \left[ \lambda_a^2k^2+ 2\lambda_b^2k^2_z \pm   2 \sqrt{(\lambda_b^2k_z^2)^2 + \lambda_a^2\lambda_b^2k_z^2k^2 } \right]
\end{gather}
This gap structure of the chiral superconductor already exhibits the generic characteristics: it is a non-unitary pairing state and has full pairing gap in the $x-y$ plane, but single non-degenerate point nodes along the $z$ axis. At these nodes one spin species remains gapless, whereas the other is gapped. 

\begin{figure}
\includegraphics[width=0.8\textwidth]{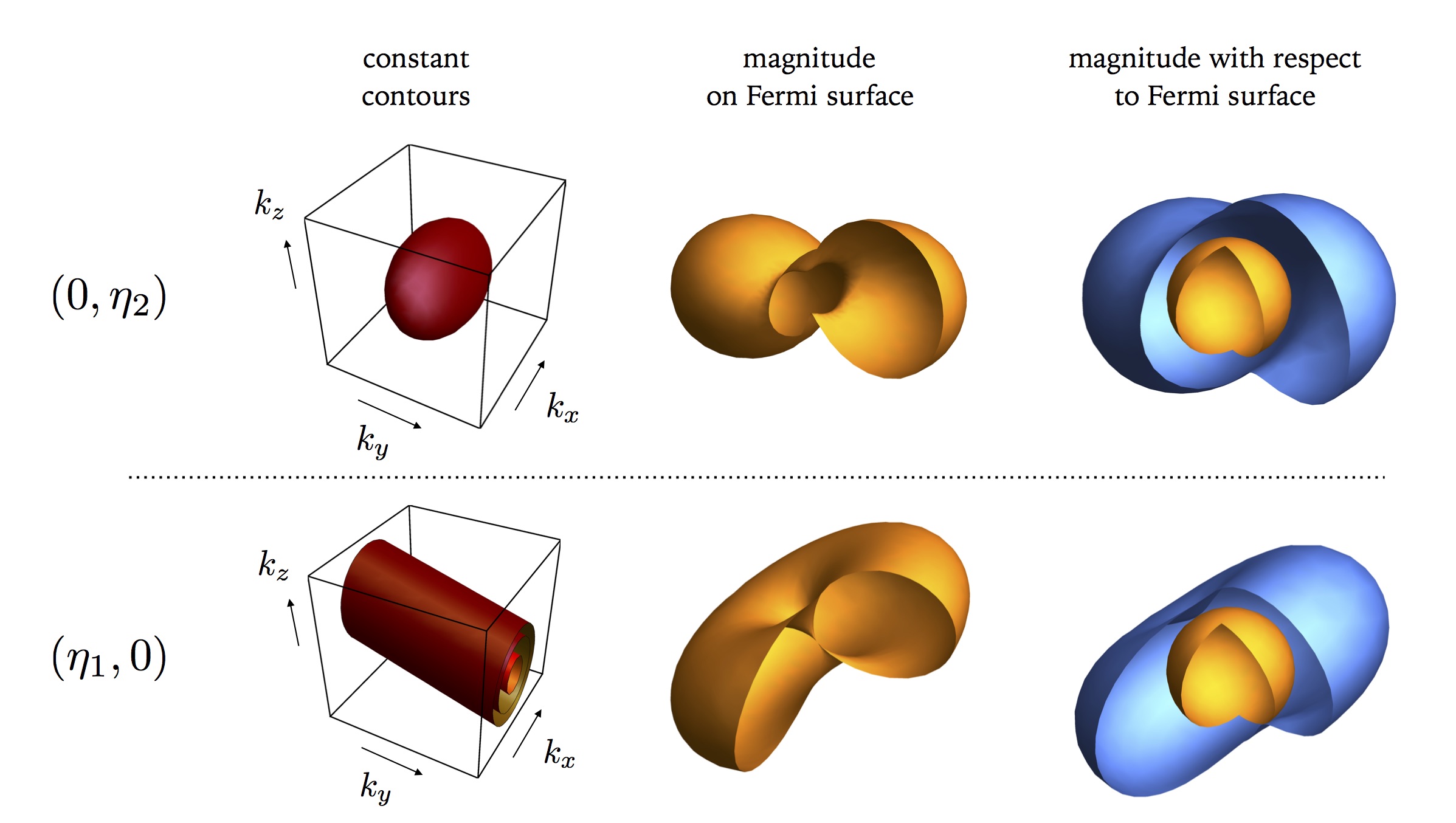}
\caption{\label{fig:gapnematic} Quasiparticle gap structure of nematic superconducting states in the $E_u$ pairing channel of trigonal symmetry. For the two superconductora $(\eta_1,0)$ and $(0,\eta_2)$ we show: (left) the surface contours of constant $ \frac{1}{2}\text{Tr}\Delta\Delta^\dagger = |d|^2$; (middle) the magnitude of $ \frac{1}{2}\text{Tr}\Delta\Delta^\dagger = |d|^2$ over the Fermi surface, i.e., as function of polar and azimuthal angle; and (right) magnitude of $ \frac{1}{2}\text{Tr}\Delta\Delta^2 = |d|^2$ (shown in blue) with respect to the Fermi surface (orange sphere) as function of polar and azimuthal angle.  }
\end{figure}

\subsubsection{Trigonal crystal symmerty: $E_{u}$ pairing}

We now turn to the trigonal crystal system and its odd-parity $E_u$ pairing channel. Taking a general linear combination of $p$-wave harmonics as before (see Eq.~\eqref{eq:pairingeu}), we have for the $(1,0)$ nematic superconductor
\begin{gather}
 \Delta_1(\vec{k}) = \eta_0 \vec d_1 \cdot \vec{\sigma}, \quad \text{with} \quad \vec d_1 =  \text{Re}\,[\lambda_a k_+ \hat{z} +  \lambda_b k_z \hat{r}_+ +  \lambda_c i k_- \hat{r}_- ] . \label{eq:nematicmirror}
\end{gather}
The pairing gap is simply obtained as
\begin{gather}
|\vec{d}|^2 = \eta^2_0 [(   \lambda_a^2 +  \lambda_c^2) k^2_x + ( \lambda_b k_z +  \lambda_c k_y )^2 ]    \label{eq:diracnematic}
\end{gather}
In the $y-z$ plane, given by $k_x=0$, this nematic superconductor has point nodes along the line $k_z = -\lambda_c k_y / \lambda_b$. In contrast, the nematic superconductor $(0,1)$ with pairing function
\begin{gather}
 \Delta_2(\vec{k}) = \eta_0 \vec d_2 \cdot \vec{\sigma}, \quad \text{with} \quad \vec d_2 = \text{Im}\,[\lambda_a k_+ \hat{z} + \lambda_b k_z \hat{r}_+ +  \lambda_c i k_- \hat{r}_- ]  \label{eq:nematicgapped}
\end{gather}
has a gap structure given by 
\begin{gather}
|\vec{d}|^2 = \eta^2_0[  \lambda_a^2k^2_y +  \lambda_c^2k^2_x + ( \lambda_b k_z -  \lambda_c k_y )^2 ] ,
\end{gather}
which corresponds to a full pairing gap. 

The difference between these two gap structures has its root in the different symmetry properties of the two superconductors. In particular, the $(1,0)$ superconductor has a mirror symmetry $M_{yz}: \; x \rightarrow - x $, which protects nodal degeneracies in the $y-z$ plane. The point nodes are twofold (i.e., spin) degenerate. In contrast, the $(0,1)$ superconductor does not have any mirror symmetry and therefore is generally fully gapped. More generally, all nematic superconductors $(\cos\theta, \sin\theta)$ which satisfy $\cos 6\theta = 1$ (i.e., are symmetry-related to the $(1,0)$ superconductor) have spin degenerate point nodes in a mirror plane. These gap structures are shown in Fig.~\ref{fig:gapnematic}.

The $ E_u$ chiral superconductor $\Delta_1(\vec{k}) + i \Delta_2(\vec{k}) $ with expansion coefficients $\lambda_{a,b,c}$ has gap structure
\begin{gather}
|\vec{d}|^2 \pm |\vec{d}^*\times \vec{d} |= \eta^2_0 \left[ ( \lambda_a^2+ 2 \lambda_c^2)k^2+ 2 \lambda_b^2k^2_z \pm   2 \sqrt{ ( \lambda_b^2k_z^2 -  \lambda_c^2 k^2)^2 +  \lambda_a^2 k^2( \lambda_b^2 k_z^2+ \lambda_c^2k^2 ) - 2 \lambda_a^2  \lambda_b  \lambda_ck_zk^3\sin 3\theta_k } \right] \label{eq:gapchiraleu}
\end{gather}
Note that the angular dependence $\sin 3\theta_k$ is consistent with three-fold rotational symmetry and the absence of time-reversal symmetry. The presence of the latter would require sixfold rotational anisotropy. Generally, the chiral superconductor is a non-unitary pairing state with \emph{single} (non-degenerate) point nodes along $z$.

\section{Dirac and Majorana nodal superconductors}

In this final part of the Supplemental Material, having discussed the gap structures of chiral and nematic superconductors in detail, we present a more detailed derivation and discussion of one of the key results: the presence of Dirac and Majorana nodes in the quasiparticle spectrum of the nematic and chiral superconductors.

\subsection{The chiral superconductor: Majorana nodes}

In order to derive an effective theory for the nodal Majorana quasiparticles we start from the chiral $E_u$ superconductor with pairing matrix
\begin{gather}
\Delta(\vec{k}) = \eta_0\Delta_1(\vec{k}) + i \eta_0\Delta_2(\vec{k}) = \eta_0 \lambda_a k_+ \sigma_z + \eta_0 \lambda_b k_z \sigma_+ + \eta_0 \lambda_c i k_- \sigma_-,
\end{gather}
describing the chiral superconductor $\eta_0(1,i)$, expanded in the three leading order degenerate basis functions with expansion coefficients $\lambda_{a,b,c}$. In addition, for simplicity we assume a spherically symmetric dispersion given by
\begin{gather}
\varepsilon(\vec{k}) = \frac{1}{2m } (k_x^2 + k_y^2 +k_z^2 ) = \frac{1}{2m } (k^2 +k_z^2 )  \equiv \varepsilon_k
\end{gather}
where we have written $\varepsilon_k$ to simplify the notation. At this stage we recall that the particle-hole Nambu spinor is defined as 
\begin{gather}
\Phi^\dagger(\vec k ) = [c^\dagger_{1}(\vec k) , c^\dagger_{2} (\vec k ) , c_{2}(-\vec k) , -c_{1} (-\vec k) ].
\end{gather}
The quasiparticle gap structure was obtained above and is given by Eq.~\eqref{eq:gapchiraleu}. The quasiparticle energy spectrum, which is given by
\begin{gather} 
\mathcal{E}_\pm(\vec{k}) = \sqrt{ (\varepsilon_k - \mu)^2 +|\vec{d}|^2 \pm |\vec{d}^*\times \vec{d} |},
\end{gather} 
with $|\vec{d}|^2 \pm |\vec{d}^*\times \vec{d} |$ given by Eq.~\eqref{eq:gapchiraleu}, vanishes for a single pseudospin species along the $z$-axis at the Fermi surface momenta $ \vec{K} = \pm (0,0,k_F)$. The other pseudospin species is gapped at $ \vec{K} = \pm (0,0,k_F)$. This is due to the non-unitary nature of the pairing state and a generic property of the chiral superconductor.

In what follows we will present the derivation for the case $\lambda_a=0$ such that the pairing decouples in (pseudo)spin space. This will be sufficient to describe the nodal Majorana quasiparticle superconductor to lowest order; we come back to the effect of $\lambda_a$ term at the end.

The superconducting mean-field Hamiltonian, expressed in the Nambu spinor basis $\Phi^\dagger(\vec k ) $, is then given by
\begin{gather}
\mathcal{H}(\vec{k}) = (\varepsilon_k -\mu ) \tau_z + \eta_0 \lambda_bk_z( \tau_x\sigma_x - \tau_y \sigma_y ) + \eta_0 \lambda_c[(k_y\tau_x - k_x\tau_y)\sigma_x +(k_x\tau_x + k_y\tau_y) \sigma_y ] \label{eq:chiralham}
\end{gather}
where we use a set of Pauli matrices $\tau_i$ to act on the (particle-hole) Nambu space and $\sigma_i$ act on spin as before. The gap structure for $\lambda_a=0$ is given by 
\begin{gather}
|\vec{d}|^2 \pm |\vec{d}^*\times \vec{d} |= \eta^2_0 \left[  2\lambda_c^2k^2+ 2\lambda_b^2 k^2_z \pm   2 (\lambda_b^2k_z^2 - \lambda_c^2 k^2) \right] 
\end{gather}
Our interest will be in the ($-$) solution, i.e., $\eta^2_0 4 \lambda_c^2k^2$, for which one pseudospin species remains gapless at $\pm \vec K$. In order to obtain an effective low-energy description for the gapless and gapped Bogoliubov quasiparticles we expand Hamiltonian \eqref{eq:chiralham} in small momenta $\vec q$ relative to two Fermi surface momenta $\pm  \vec K$. We define the corresponding Nambu spinors $\Phi_\pm(\vec q )$ as
\begin{gather}
\Phi_+(\vec q ) = \Phi(\vec K + \vec q ) = \begin{pmatrix} c_{1}(\vec K+\vec q)  \\ c_{2}(\vec K+\vec q)  \\ c^\dagger_{2}(-\vec K -\vec{q}) \\ -c^\dagger_{1}(-\vec K-\vec q)  \end{pmatrix}, \quad \Phi_-(\vec q ) = \Phi(-\vec K + \vec q ) = \begin{pmatrix} c_{1}(-\vec K+\vec q)  \\ c_{2}(-\vec K+\vec q)  \\ c^\dagger_{2}(\vec K -\vec{q}) \\ -c^\dagger_{1}(\vec K-\vec q)  \end{pmatrix}, \label{eq:nambuexpand}
\end{gather}
In terms of these Nambu spinor operators the Hamiltonian in the vicinity of $\pm \vec K$, expanded up to linear order in $\vec q$, can be written as 
\begin{gather}
H = \frac{1}{2}\sum_{ n=\pm} \sum_{\vec q } \Phi^\dagger_n(\vec q ) \mathcal{H}_n(\vec{q})  \Phi_n(\vec q ), 
\end{gather}
with the Hamiltonians $ \mathcal{H}_\pm(\vec{q})$ of the $\pm $ blocks given by
\begin{gather}
 \mathcal{H}_\pm(\vec{q})   =   \begin{pmatrix}  \pm v_F q_z & 0 & 0& \pm 2\eta_0\lambda_b(k_F \pm q_z) \\ 0 & \pm v_F q_z& 2\eta_0\lambda_c i q_-& \\ 0& - 2\eta_0\lambda_c iq_+ & \mp v_F q_z &\\ \pm 2\eta_0\lambda_b(k_F \pm q_z) & 0&0& \mp v_F q_z\end{pmatrix}  . \label{eq:hampm}
\end{gather}
Here $v_F$ is the Fermi velocity and originates from expanding $\varepsilon_k - \mu$. The next step is to separate the gapped and gapless degrees of freedom by performing the appropriate basis transformation in the space of $(\Phi_+,\Phi_-)$. From Eq.~\eqref{eq:hampm} it is straightforward to find such a basis by rearranging the electron and hole operators according to the pseudospin label $\sigma_z = \pm 1$. We collect the $\sigma_z = \pm 1$ operators in new spinors $\Psi_{1,2}(\vec q ) $, which are given by
\begin{gather}
\Psi_1(\vec q ) =  \begin{pmatrix} c_{1}(\vec K+\vec q)  \\ c_{1}(-\vec K+\vec q)  \\ c^\dagger_{1}(\vec K -\vec{q}) \\ c^\dagger_{1}(-\vec K-\vec q)  \end{pmatrix}, \quad \Psi_2(\vec q ) =  \begin{pmatrix} c_{2}(\vec K+\vec q)  \\ c_{2}(-\vec K+\vec q)  \\ c^\dagger_{2}(\vec K -\vec{q}) \\ c^\dagger_{2}(-\vec K-\vec q)  \end{pmatrix}. \label{eq:majorana}
\end{gather}
In terms of these new operators the Hamiltonian can be written as $H= H_1 + H_2 $, with the first term $H_{1}$ given by (ignoring the linear in $q_z$ contribution from the pairing)
\begin{gather}
H_1 = \frac{1}{2} \sum_{\vec q } \Psi^\dagger_1(\vec q )    \begin{pmatrix}   v_F q_z & 0 & 0& -2\eta_0\lambda_b k_F \\ 0 & - v_F q_z& 2\eta_0\lambda_b k_F& \\ 0& 2\eta_0\lambda_b k_F &  v_F q_z &\\  -2\eta_0\lambda_b k_F & 0&0& - v_F q_z\end{pmatrix}  \Psi_1(\vec q )  .
\end{gather}
It is convenient to introduce a new set Pauli matrices corresponding to the label $\pm \vec K$, and we denote the set Pauli matrices by $\mu_i$. Then, Hamiltonian $\mathcal{H}_1(\vec q)$ takes the simple form $\mathcal{H}_1(\vec q) = v_F q_z \mu_z +m \tau_y\mu_y$ with $m = 2\eta_0\lambda_b k_F$. The mass term $m \tau_y\mu_y$, which is proportional to $k_F$, implies that the gapped quasiparticles are given by the field $\Psi_1(\vec q )$ and correspond to spin $\sigma_z = +1$. In contrast, the low-energy Hamiltonian for the $\Psi_2 (\vec q )$ quasiparticles is given by
\begin{gather}
H_2 = \frac{1}{2} \sum_{\vec q } \Psi^\dagger_2(\vec q ) \begin{pmatrix}  v_F q_z & 0 & 0& 2\eta_0\lambda_c iq_- \\ 0 &-v_F q_z& 2\eta_0 \lambda_c iq_-& \\ 0& -2\eta_0\lambda_c iq_+ & v_F q_z &\\  -2\eta_0\lambda_c iq_+& 0&0& -v_F q_z\end{pmatrix}   \Psi_2 (\vec q ), \label{eq:majorananode}
\end{gather}
and describes massless quasiparticles. This is the Hamiltonian quoted in the main text. As before, we use the Pauli matrices $\mu_i$ to describe the $\pm \vec K$ degree of freedom and write the Hamiltonian as $\mathcal{H}_2(\vec q) = v_F q_z \mu_z - v_\Delta \mu_x ( q_x \tau_y - q_y\tau_x)$, where $v_\Delta =  2\eta_0\lambda_c $. To cast this into an even more transparent form, we use the rescaled momenta $\tilde{q} = (v_\Delta q_x,v_\Delta q_y,v_Fq_z)$ and perform a basis change by switching the hole operators ($ c^\dagger_{2}(\vec K -\vec{q}) \leftrightarrow c^\dagger_{2}(-\vec K-\vec q) $). The Hamiltonian is then expressed as
\begin{gather}
\mathcal{H}_2(\tilde{q}) =\tilde{q}_z \tau_z\mu_z  - \tilde{q}_x \tau_y + \tilde{q}_y\tau_x ,
\end{gather}
which has the structure of a massless Dirac Hamiltonian in particle-hole space $\tau_i$ with an additional grading $\mu_z$. The additional grading, corresponding to the two nodes at $\pm \vec K$, gives the two nodes opposite chirality. We now show that the low-energy quasiparticles are not complex Dirac fermions, but real Majorana fermions. 

\begin{figure}
\includegraphics[width=0.8\textwidth]{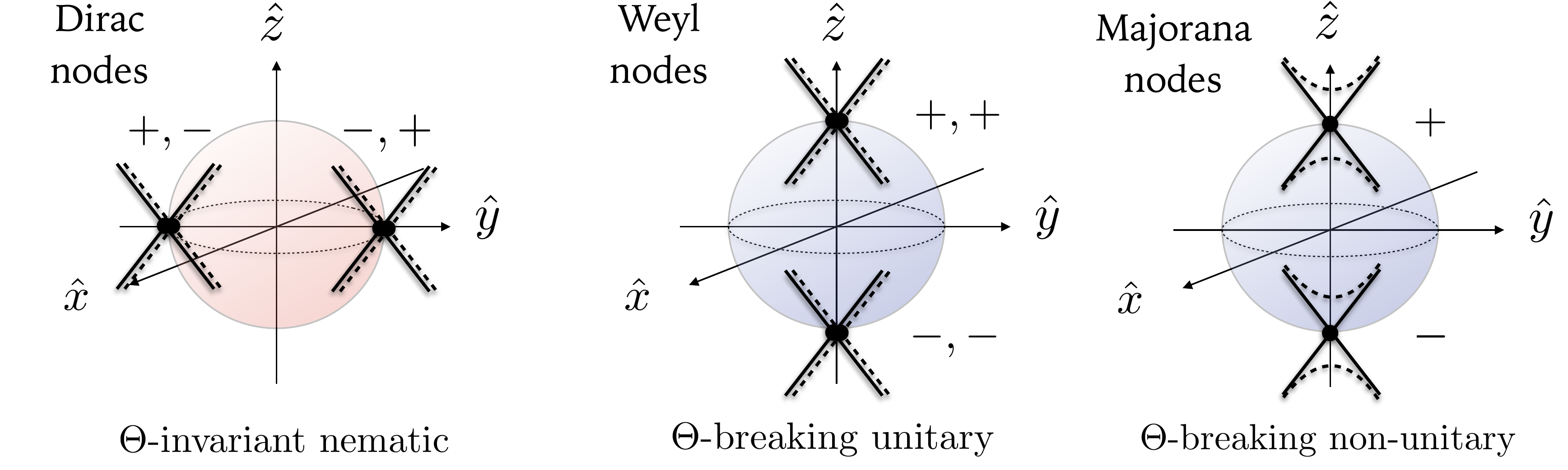}
\caption{\label{fig:pointnodes} Cartoon picture of the point nodal quasiparticle gap structure of $\Theta$-invariant nematic superconductors with mirror symmetry, $\Theta$-breaking but unitary chiral superconductors, and $\Theta$-breaking non-unitary chiral superconductors. An example of a nematic superconductor with spin-degenerate Dirac nodes on the $y$ axis is shown on the left. The two degenerate nodes have opposite chirality ($\pm$). In case of the chiral unitary superconductor, which is realized in $A$-phase of superfluid $^3$He, the degenerate Weyl point nodes have the same chirality, but the $\pm \vec K$ nodes have opposite chirality. The non-unitary chiral superconductor has \emph{non}-degenerate Majorana point nodes and a branch of gapped quasiparticle excitations. The nodal structure is reflected in the surface state excitation structure: the nematic superconductor hosts Majorana Kramers arcs, the chiral unitary superconductor hosts chiral Fermi arcs, and the chiral non-unitary superconductor hosts chiral Majorana arcs. }
\end{figure}

We define the four-component quantum fields $\Psi_i(\vec x )$ by taking the Fourier transform of $\Psi_i(\vec q )$, i.e.,
$ \Psi_i(\vec x )  = \sum_{\vec q} e^{-i\vec q \cdot \vec x} \Psi_i(\vec q )$. From Eq.~\eqref{eq:majorana} we then deduce that both fields $\Psi_i(\vec x )$ satisfy a reality condition, which takes the following form
\begin{gather}
\Psi^\dagger_i(\vec x )  = (\tau_x \Psi_i(\vec x ) )^T.
\end{gather}
This reality condition satisfied by the four-component quasiparticle field is the defining property of four-component real Majorana fermions. Each of the two fields $\Psi_{1,2}(\vec x )$ is therefore a realization of the Majorana fermion field known in particle physics. In general, two flavors of Majorana fermion fields can be used to define a complex Dirac fermion field as
\begin{gather}
\Upsilon(\vec x )   =  \Psi_1(\vec x ) + i \Psi_2(\vec x )
\end{gather}
Since in our case one of the flavors of Majorana fermions is gapped, the low-energy quasiparticles are described by a \emph{single} Majorana fermion field. As a result, the quasiparticle spectrum of odd-parity chiral superconductors shows Majorana point nodes. A cartoon picture of the Majorana nodes is shown in Fig.~\ref{fig:pointnodes}.

At this stage, it is worth commenting on the term $ k_+ \sigma_z $. As we mentioned, explicitly including it in the derivation of the Majorana nodes does not change Hamiltonian~\eqref{eq:majorananode}. In particular, including the coupling of (pseudo)spin species implied by $ k_+ \sigma_z $ generates contributions which are of higher order in momentum $\vec q$. It is illuminating, however, to consider a different starting point by choosing the chiral pairing $\Delta(\vec{k}) = \eta_0 \lambda_a k_+ \sigma_z $, i.e., taking $\lambda_b=\lambda_c=0$. Then $\Delta(\vec{k}) $ is equivalent to the order parameter of the $A$-phase of superfluid $^3$He~\cite{volovik_SM,salomaa_SM,silaev_SM}. In particular, the quasiparticle gap structure has spin-degenerate point nodes along the $z$ axis at $\pm \vec K$. The Hamiltonian governing the low-energy quasiparticles is simply obtained as $ \mathcal{H}_\pm(\vec{q})   =   \pm v_F q_z  \tau_z + \overline{v} (q_x\tau_x - q_y \tau_y  )\sigma_z   $, where $\overline{v} =  \eta_0 \lambda_a$. This implies that the degenerate nodes at $+\vec K$ ($-\vec K$) both have positive (negative) chirality. As a result, the nodal structure is referred to as (Bogoliubov-)Weyl points~\cite{volovik_SM,balents_SM}. A cartoon picture of the Weyl nodes is shown in Fig.~\ref{fig:pointnodes}. 

In an important paper by Meng and Balents~\cite{balents_SM}, the concept of Weyl semimetals was generalized to Bogoliubov-Weyl superconductors, and it was shown that the degenerate Weyl nodes can be split, resulting in non-degenerate point nodes of the kind we identified as Majorana nodes. Similarly, non-spin-degenerate point nodes in the Bogoliubov quasiparticle spectrum are present in ferromagnetic triplet superconductors, where the spin-degeneracy is lifted due to ferromagnetism~\cite{sau_SM}. The single Majorana nodes of the chiral superconductor presented in this work can be understood as gapping out one branch of spin-polarized nodes, with the other branch remaining gapless.


\subsection{The nematic superconductor: Dirac nodes}

We conclude this section with a discussion of the (spin-)degenerate point nodes in the quasiparticle spectrum of nematic superconductors with mirror symmetry. The low-energy description of the gapless quasiparticles can be obtained in a straightforward way. We recall from Eq.~\eqref{eq:nematicmirror} that the pairing matrix of the superconductor $(\eta_1,\eta_2) = \eta_0 (1,0 )$, which possesses the mirror symmetry $M_{yz}: \; x \rightarrow  - x$, is given by
\begin{gather}
 \Delta(\vec{k}) = \eta_0[ \lambda_ak_x\sigma_z + \lambda_bk_z\sigma_x  + \lambda_c( k_x\sigma_y +k_y\sigma_x ) ].
\end{gather}
To illustrate the basis features we consider the case $ \lambda_c = 0$. In this case the corresponding gap structure, given in Eq.~\eqref{eq:diracnematic}, exhibits point nodes along the $y$ axis at $\vec K  = (0,\pm k_F , 0)$. Expanding the Hamiltonian in small momenta $\vec q $ relative to these gapless points, in the same way as above, one directly obtains the Hamiltonian
\begin{gather}
 \mathcal{H}_\pm(\vec{q})   = \pm v_F q_y \tau_z +  \eta_0\lambda_a q_x \tau_x \sigma_z +  \eta_0\lambda_b q_z \tau_x \sigma_x,\label{eq:hamdirac}
\end{gather}
expressed in terms of the basis $(\Phi_+,\Phi_-)$, which is defined in the same way as in Eq.~\eqref{eq:nambuexpand}, but with $\vec K  = (0,\pm k_F , 0)$. Rescaling the momenta $\vec q$ as $\tilde{q} = (\eta_0\lambda_a q_x, v_F q_y , \eta_0\lambda_b q_z ) $ the Hamiltonian can be reexpressed as
\begin{gather}
 \mathcal{H}_\pm(\tilde{q})   = \pm \tilde{q}_{y} \tau_z + \tau_x (  \tilde{q}_{x} \sigma_z +  \tilde{q}_{z}  \sigma_x), \label{eq:diracnematicLE}
\end{gather}
from which it is clear that it has Dirac structure. Due to the pseudospin degeneracy of the nodes (i.e., both $ \mathcal{H}_+$ and $\mathcal{H}_-$ are gapless), the low-energy quasiparticles are described by a complex Dirac field. A cartoon picture of the nodal structure is shown in Fig.~\ref{fig:pointnodes}.

Introducing an additional Pauli matrix grading $\nu_z = \pm 1$ to describe the $\pm \vec K$ degree of freedom, the Hamiltonian takes the form
\begin{gather}
 \mathcal{H}(\vec{q})   =  \tilde{q}_{y} \nu_z \tau_z  + \tau_x (  \tilde{q}_{x} \sigma_z +  \tilde{q}_{z}  \sigma_x). 
\end{gather}
The mass term $\nu_z\tau_x\sigma_y$, which is parity odd, time-reversal invariant and anti-commutes with $ \nu_z \tau_z $, $ \tau_x  \sigma_z$, and $\tau_x  \sigma_x$, violates the mirror symmetry $M_{yz}$ and is therefore symmetry-forbidden~\cite{zhang_SM}. This would be different for the nematic superconductor $(0,1)$, where one can develop a low-energy point nodal theory for the nodes along the $x$ axis (again assuming $\lambda_c=0$). In that case, the appropriate mass term, given by $\nu_z\tau_x\sigma_x$, does not break any additional symmetries and is therefore allowed. It is generated by mixing in of $\lambda_c$. Ref.~\onlinecite{zhang} has discussed such symmetry breaking perturbations, including the breaking of time-reversal symmetry, which lifts the degeneracy of the point nodes and leads to a doubled but non-degenerate set of Majorana nodes (see above).

Going back again to Hamiltonian~\eqref{eq:diracnematicLE} we can perform a rotation in spin space to arrive at the following expression for the Hamiltonian
\begin{gather}
 \mathcal{H}_\pm(\vec{q})   = \pm \tilde{q}_{y} \tau_z + |\tilde{q}_{\perp}| \tau_x \sigma_z  ,
\end{gather}
where $|\tilde{q}_{\perp}|$ is the rescaled momentum in the plane perpendicular to $\tilde{q}_{y}$. Observe that the spin label $\sigma_z$ now provides a grading for the low-energy quasiparticle dispersion, giving the degenerate nodes at $\vec K$ (and $-\vec K$) opposite chirality. This is consistent with time-reversal symmetry and an effective inversion symmetry of the Bogoliubov-de-Gennes Hamiltonian. In addition, it is related to the Majorana Kramers arcs on the surface of the nodal superconductor, which were derived in Ref.~\onlinecite{zhang}.

In the general case, when $ \lambda_c \neq 0$, the location of the nodes simply moves in the $y-z$ plane, $\vec K  = \pm k_F (0, \cos \zeta , \sin \zeta)$. The nodes are required to be pseudospin degenerate due to time-reversal symmetry.

%
%
%


\bibliographystyle{apsrev}

\begin{thebibliography}{10}
\bibitem{note1} If spin-orbit interaction is adiabatically switched off, the MCBB smoothly evolve into spin eigenstates, i.e., $c_{1} \rightarrow c_{\up}$ and $c_{2} \rightarrow c_{\down}$.
\end{thebibliography}

\end{document}